\documentclass[num-refs]{wiley-article}

\usepackage{siunitx}
\usepackage{hyperref}
\hypersetup{
    colorlinks=true,
    linkcolor=blue,
    urlcolor=blue,
    citecolor=cyan,
    pdftitle={Verleyen_2022_Framework_AI_ML_Innovation},
    pdfpagemode=FullScreen,
    }

\papertype{arXiv - cs: computers and society}

\title{Framework for disruptive AI/ML Innovation}

\author[1\authfn{1}]{Wim Verleyen PhD}
\author[1]{William McGinnis}

\affil[1]{Enterprise Data Services, Raytheon Technologies, Brooklyn, NY, 11207, US}

\corraddress{Wim Verleyen PhD, Head of AI/ML Innovation, Enterprise Data Services, Raytheon Technologies, Brooklyn, NY, 11201, US}
\corremail{wim.verleyen@rtx.com}

\presentadd[\authfn{2}]{Enterprise Data Services, Raytheon Technologies, Brooklyn, NY, 11201, US}

\runningauthor{Verleyen, W. et al.}

\begin{document}

\begin{frontmatter}
\maketitle

\begin{abstract}
\texttt{Abstract}

The revenue forecast for 2022 for worldwide AI software is \$62.5 billion (an increment of 21.3\% compared to 2021); by 2030, recent market research estimates that AI will represent \$13 trillion in value creation. Therefore, a framework for disruptive AI/ML Innovation (AMI) has never been more prominent. This framework enables C-suite executive leaders to define a business plan and manage a set of technological dependencies for building AI/ML Solutions.

The business plan of this framework provides components and background information to define strategy and analyze cost; furthermore, the business plan represents the fundamentals of AMI and AI/ML Solutions. Therefore, the framework provides a menu for managing and investing in AI/ML.

Finally, this framework is constructed with an interdisciplinary and holistic view of AMI and builds on advances in business strategy in harmony with technological progress for AI/ML. This framework incorporates an organization's value chains, supply chain, and ecosystem strategies. 

\end{abstract}
\end{frontmatter}

\section{Introduction}
\label{section:introduction}

\subparagraph{Current business impact of AI/ML} One of the main drivers of interest in AI/ML is to supplement humans on important tasks within an organization \cite{agrawal2017expect,daugherty2018human,amabile2020creativity,de2021ai,top2021ai}, e.g., conversational AI chatbots \cite{gao2018neural}, language translation \cite{vaswani2017attention}, materials discovery \cite{tran2018active}, pharmaceutical -and biomedicine research \cite{schuhmacher2020upside,wainberg2018deep,mamoshina2016applications, stokes2020deep}, breast cancer classification \cite{wu2019deep,mckinney2020international}, protein structure prediction and protein engineering \cite{yang2019machine,jumper2021highly}, etc. Recent developments have illustrated that AI/ML outperforms humans, e.g., playing Go \cite{silver2017mastering}, designing floorplans for microchips \cite{kahng2021ai}, tumor classification for precision medicine \cite{wu2021radiological}, AI-generated art \cite{ragot2020ai}, etc.   Despite the progress in AI/ML, there are still major limitations \cite{chui2018ai,hagendorff202015}, i.e., the dependency on data, a lack of emotional intelligence, the known challenges related to AI safety, the limitations related to data privacy and AI regulations, etc. One of the main reasons to define this framework is the potential lack of business gains from AI-related investments \cite{davenport2018artificial}; recently, a global study showed that only 10\% of the organizations see a financial benefit from investing in AI \cite{ransbotham2020expanding}. The disruptive nature and the dependency on a set of technologies of AI/ML Innovation (AMI) imposes challenges on the entire organization.

\begin{figure}[!ht]
\centering
\includegraphics[scale=0.94]{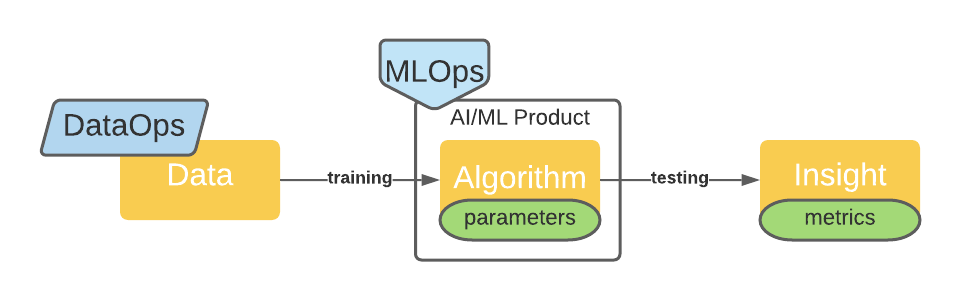}
\caption{The experimental design during AI/ML Product development}
\label{fig:AI_ML_product_development}
\end{figure}

\subparagraph{AI/ML Product} More recently, a vital abstraction, an AI/ML Product (see Figure~\ref{fig:AI_ML_product_development}), helps the overall management of work; it allows us to separate the algorithm development. Therefore, the AI/ML Product is the algorithmic solution that can be re-used and evaluated across multiple use cases. Therefore, we can concentrate on the core computational innovation and define a product portfolio for AI/ML.

\subparagraph{Experimental design} An essential concept for performing effective AI/ML Product development is the experimental design encapsulating the training and testing of the AI/ML Product. This experimental design is critical during the development cycle. Practical experimentation enables researchers to save time and cost during AI/ML Product development \cite{robertson2021cloud}. An experimental design represents three main steps (see Figure~\ref{fig:AI_ML_product_development}): (1) data: in order to perform practical experimentation and training of the algorithm, the quality of the data is fundamental (often referred to as garbage-in garbage-out) \cite{gigo2019}, (2) AI/ML Product: the algorithm learns the optimal parameters during training, and (3) insight: the prediction that enables the business practitioners to make better decisions and evaluated by performance metrics. This experimental design can evolve into the AI/ML Benchmark of the corresponding AI/ML Product. This AI/ML Benchmark will evaluate the performance of an AI/ML Product during a deployment with the customer and compare its performance before and after AI/ML Product improvements. Finally, these AI/ML Benchmarks facilitate further innovation by AI/ML practitioners within an organization.

\subparagraph{Software solution architecture} The experimental design often depends on the software solution architecture in an organization. This architecture will contain two generic sets of technologies: (1) DataOps \cite{ereth2018dataops,munappy2020ad}, and (2) MLOps \cite{tamburri2020sustainable,makinen2021needs}. DataOps is a set of technologies that support data governance, data verification, data ingestion, data storage, data catalogs, data preprocessing, etc. DataOps can additionally cover procedures and protocols to guarantee data updates, data quality, backup procedures, continuous delivery, etc. MLOps is a set of practices and technologies that support machine learning and deep learning for AI/ML Product development and deployment, i.e., feature store, feature selection, model verification, etc. The deployment of an AI/ML Product reproduces the experimental design with additional requirements related to the quality assurance of the produced insight. The development of these DataOps and MLOps solutions follow the Agile methodology \cite{robertson2021cloud,highsmith2009agile}. Since the solution architectures for MLOps are getting more established, there is a trend to build reusable elements with so-called design patterns \cite{lakshmanan2020machine}. Similarly, the concept of design patterns improved the software architectures of object-oriented implementations \cite{gamma1995elements}. Additionally, best practices of DevOps in an organization should enable continuous integration and continuous delivery (CI/CD), including MLOps and DataOps capabilities.

\section{Innovation}
\label{section:innovation_frameworks}

\begin{figure}[!ht]
\centering
\includegraphics[scale=0.53]{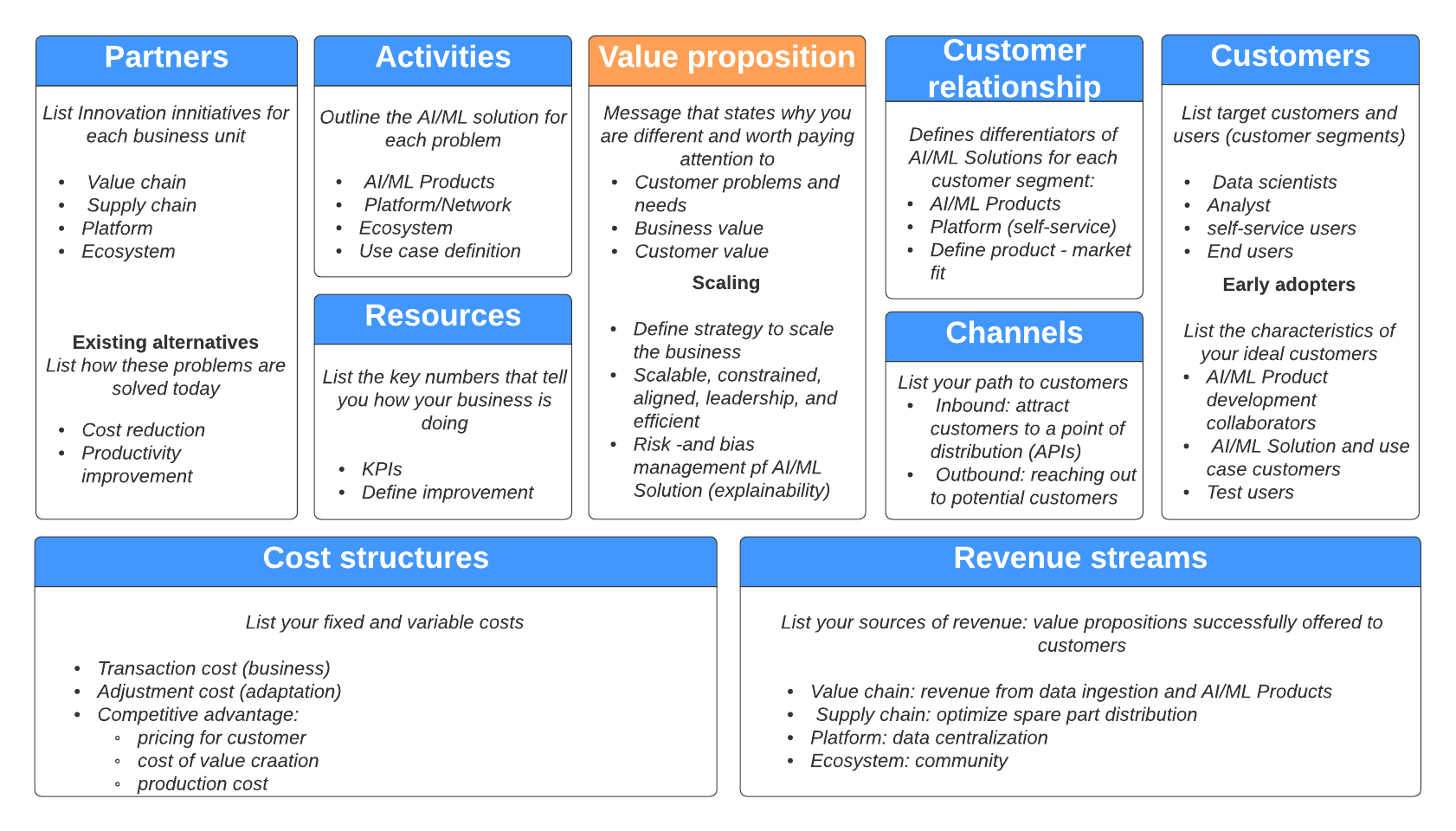}
\caption{The business model canvas for AI/ML Innovation (AMI) initiative.}
\label{fig:AI_ML_business_model_canvas}
\end{figure}

\subparagraph{Managing innovation} Innovation management needs to address three main levels: (1) the strategic level, (2) the process and pipeline level, and (3) the project level. Innovation is the main activity of an organization for creating economic value in the market \cite{price200612}. Innovation is not only about ideas and inventions; concretely, innovation defines a solution addressing a need of customers in the market. Innovation needs to define business opportunities connecting organizational -and business strategy from the C-level leadership with embracing culture for ambiguity and delivery. Implementing a process of innovation can be very uncertain; we need to translate an idea or invention into a market opportunity and a business plan to generate revenue. Generally, there are three main stages during an innovation process: (1) the generation of ideas, (2) the development of ideas, and (3) the evaluation and selection of most promising ideas \cite{kijkuit2007organizational}. Innovation initiatives and projects should reflect in the business model. A business model defines all parts of commercialization that create value for the customers and the organization \cite{osterwalder2010business,blank2013lean}. A business model consists of 9 building blocks \cite{osterwalder2010business}: (1) partners, (2) activities, (3) resources, (4) value proposition, (5) customer relationship, (6) channels, (7) customers, (8) cost structures, and (9) revenue streams (see Figure~\ref{fig:AI_ML_business_model_canvas}). Generating a business model is dependent on the four levels of innovation: (1) \emph{incremental innovation} results in gradual improvement to existing products and technologies \cite{henderson1990architectural,richter2018entrepreneurial}, (2) \emph{sustaining innovation} results in incremental changes of a product or adding small features for satisfying existing customers \cite{bower1996disruptive}, (3) \emph{radical innovation} results in product development for new markets with new technologies for the creation of novel capabilities \cite{tripsas2000capabilities,ahuja2001entrepreneurship}, and (4) \emph{disruptive innovation} results in fundamental novel products or services based upon novel technologies with business model innovation \cite{christensen2013innovator}. It is important to note that disruptive innovation will result in a business model with much more hypotheses rather than facts \cite{blank2016}. Often, disruptive innovation happens in startups entering a market with emergent technology that is not on incumbents' radar. Innovation can occur in different forms: (1) innovation as something new, (2) innovation as a conduit to change, (3) innovation as a process, (4) innovation as a value driver, and (5) innovation as invention \cite{price200612}.

\subparagraph{Innovation frameworks} The three horizons framework often reviews innovation \cite{baghai2000alchemy}. Typically, each horizon characterizes two dimensions: (1) the level of novelty of the market and its customers related to a business model and (2) the novelty of the technology or solution. In horizon 1, innovation initiatives in mature businesses are performed in a well-known market with well-understood technology by the organization. The organization can execute an existing business model. The business plan formulates a roadmap of activities between 0 and 18 months, i.e., extend features on an existing product, improve supply chain efficiency, change pricing, etc. Many organizations spend 50-70\% of their total innovation budget on horizon 1. This innovation often occurs in a permanent organization designed to execute a repeatable and scalable process. In horizon 2, the innovation initiatives in fast-growing businesses operate in a new market and technology for the organization. Often the business model needs to be extended, i.e., adding a channel, targeting a different customer segment with the same product, developing a new product for an existing channel and customer segment, etc. The business plan covers a period between 12 and 36 months, and it represents 20-30\% of the total innovation budget. In horizon 3, an innovation initiative in an emergent business is performed in a new to the world market with new technology. The organization is searching for a new business model. The business plan for a horizon 3 initiative formulated a period between 36 and 72 months. Many organizations define an innovation portfolio instead of single innovation initiatives to mitigate the high risk and the long-term return-of-innovation. Often, this innovation happens in a temporary organization designed to search for a repeatable and scalable process. This framework helps to evaluate the risk of innovation. Typically, most organizations try to avoid horizon 3 because of the associated uncertainty.

\begin{figure}[!ht]
\centering
\includegraphics[scale=0.55]{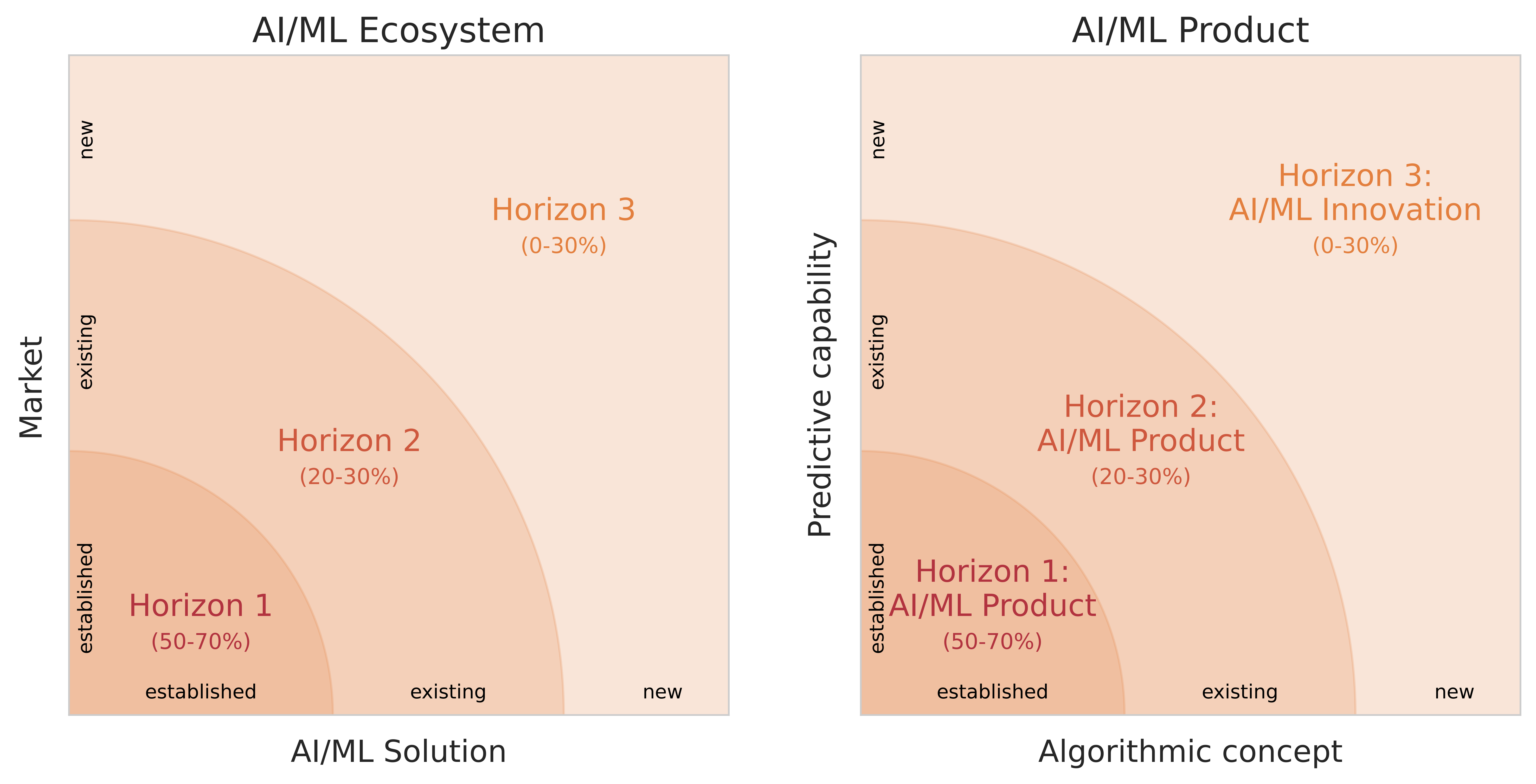}
\caption{The three horizons framework for AI/ML Innovation (AMI).}
\label{fig:AI_ML_horizons}
\end{figure}

\subparagraph{AI/ML Innovation (AMI)} Since AMI has multiple dependencies, we propose to decouple the AI/ML Product development from the AI/ML Solution. Therefore, we introduce a second three horizons framework related to the AI/ML Product. Each horizon for a product represents two dimensions: (1) the \emph{predictive capability} resulted in the AI/ML Product and (2) the \emph{algorithmic concepts} applied in the AI/ML Product. In horizon 1, the capability performs predictions with existing algorithm concepts. Now, we need to execute the experimental design to evaluate the performance of the AI/ML Product. The outcome of the performance evaluation should represent the likelihood of success of the innovation initiative; this success is mainly dependent on the data quality (50-70\%; 0-6 months). In horizon 2, the capability and/or the algorithmic concept needs an extension in order to be able to apply the AI/ML Product. Potentially, a horizon 1 innovation can lead to a horizon 2 innovation if the experimental design indicates shortcomings of the AI/ML Products. A horizon 2 effort can demand an extension of the experimental design to evaluate a novel predictive capacity (20-30\%; 6-24 months). In horizon 3, a new AI/ML Product is developed; this development requires novel algorithmic concepts. 

In practice, AMI innovation requires adjustments in the business model and incorporates novel technologies. The process of innovation is challenging. Managing innovation with new AI initiatives is important to improve the chances of a successful outcome. Innovation based upon a combination of existing technologies could overcome the lack of the current maturity of AI-related technologies and standards \cite{wu2022}. Traditionally, innovation is often studied in the context of an entering firm or startup disrupting the market with emergent technology. This market has an incumbent leader \cite{henderson1990architectural,foster1986innovation}. More recently, as technological solutions become more and more complex, many incumbent organizations are more focused on internal innovation. More organizations introduce incubators or organize hackathons or datathons to stimulate innovative initiatives. Therefore, a framework channeling disruptive AMI initiatives will still incorporate and build on existing innovation frameworks and strategies. Recently, the lean methodology introduced a guide to the difference between a startup and incumbent organization \cite{blank2013lean}. This methodology focuses on a startup, and it is composed of 3 main parts: (1) a business model canvas to define the hypotheses important for defining the future business model \cite{osterwalder2010business}, (2) customer development by designing a set of experiments for answering the hypotheses from the business model canvas in order to define the value proposition for your customers, and (3) agile engineering for incrementally and iteratively delivering a minimal viable product \cite{ries2014lean}.

\section{Organizational strategy}
\label{section:organizational_strategy}

\begin{figure}[!ht]
\centering
\includegraphics[scale=0.3]{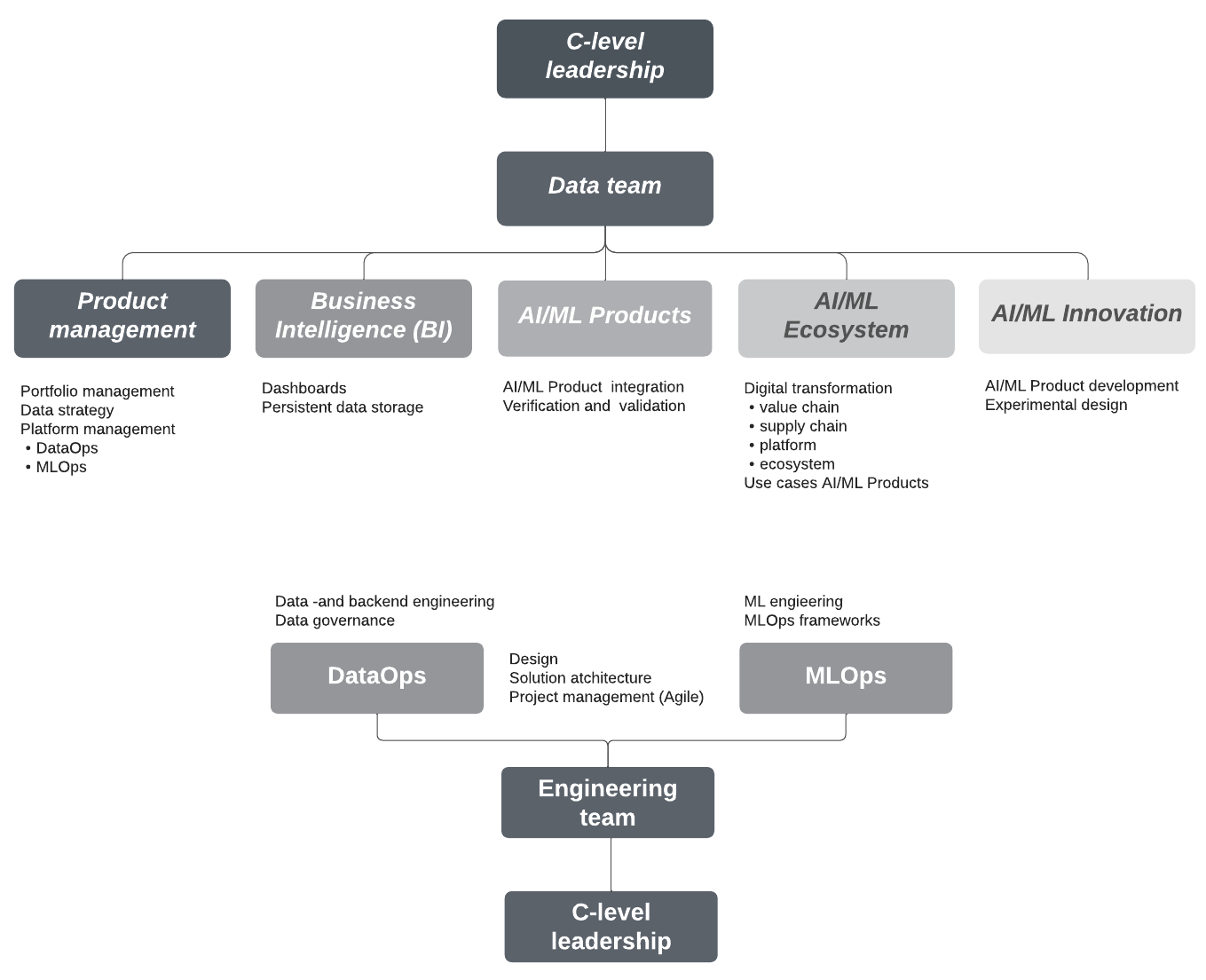}
\caption{The organigram of the Data -and Engineering teams related to the AI/ML organization strategy.}
\label{fig:AI_ML_organigram}
\end{figure}

\subparagraph{Data team} In order to enable AI/ML capabilities, the organizational strategy is crucial to define clear roles and responsibilities \cite{nyc2021aistrategy}. This section will address the organizational strategy for the data team; this team will not be responsible for any engineering activities outside of the experimental design and the deployment of an AI/ML Product (see Figure~\ref{fig:AI_ML_organigram}). Here, we will propose the internal organization for a data team. Further research suggests an organizational strategy based upon three parts \cite{christensen2000meeting}: (1) an internal innovation and product development, (2) an external spin-out, or (3) a merge and acquisition for incorporating key capabilities into the AI/ML Product portfolio. Additionally, innovation efforts can be centralized, decentralized, or a hybrid approach \cite{borchert2003operatives}. AI/ML requires internal innovation and AI/ML Product development. An AI/ML capability can be purchased; often, such an AI/ML firm's primary value resides in their workforce. Therefore, a strategy must be defined for adaptation of this AI/ML acquisition. Alternatively, a novel AI/ML capability can become a central part of an organization and could fulfill a central need for customers. Finally, an organization can decide to build an external spin-out. Generally, the nature of AMI will focus on a centralized team concentrating on building internal capabilities. 

Let us start with the analytic functions of the data team. Potentially, there are four different functions that need to be covered (see Figure~\ref{fig:AI_ML_organigram}): (1) business intelligence (BI), (2) AI/ML Ecosystem, (3) AI/ML Product, and (4) AI/ML Innovation (AMI). The BI function provides dashboards. These dashboards report the summary statistics derived from data important for the business. BI encapsulates two important types of technology: (1) persistent data storage and (2) a dashboard. Depending on the organization's data maturity, it can be beneficial to start with a BI initiative instead of an AMI initiative. These BI capabilities provide a good starting point for future extension with predictive analytics based upon AI/ML.

\subparagraph{AI/ML Ecosystem function} The AI/ML Ecosystem function supports the digital transformation related to AI/ML. This digital transformation could impact the supply chain, the value chain, the platform, and the ecosystem business practices. It is a crucial function to keep track of the disruptive nature of the AI/ML Products with corresponding use cases. Identifying potential bottlenecks, the request for proposals (RFP) for new use cases for AI/ML Products, community management, etc., in the ecosystem are essential topics to manage.

\subparagraph{AI/ML Innovation (AMI) function} The AMI function is responsible for new AI/ML Product development. This AI/ML Product development embodies radical -and disruptive innovation. The data team, the AMI function, the product management function, an executive sponsor, a customer, etc., can initiate a new AI/ML Product development. It is a crucial priority to enable the team with the appropriate tools for practical experimentation. The experimental design and its implementation are crucial for finding the dominant design of the AI/ML Product.

\subparagraph{AI/ML Product function} The AI/ML Product function is responsible for adopting new AI/ML Products into the platform-based ecosystem. Additionally, the AI/ML Product team performs incremental -and sustaining innovation, i.e., maintaining the AI/ML Product for deployment by validation and verification procedures, extending the support of the AI/ML Product across the set of hosting environments, integrating the AI/ML Product into platforms, performing small AI/ML Product development efforts, and circling back on new AMI initiatives related to existing AI/ML Products.

\subparagraph{MLOps and DataOps} Furthermore, the data team needs sufficient resources to support DataOps and MLOps functions. DataOps contains design, project management, product management, solutions architecture, data and backend engineering, data governance, and platform engineering talent. Similarly, MLOps contains design, project management, product management, solutions architecture, ML engineering, and MLOps framework engineering talent. The DataOps and MLOps functions can be overlooked or not explicitly assigned to the data team. It is crucial to budget for DataOps and MLOps to establish a data culture in an organization. Some of these DataOps and MLOps solutions will be part of a platform. Such a platform can leverage an organization's on-premise, cloud, or hybrid infrastructure. Furthermore, these DataOps and MLOps capabilities will allow for collaboration between AI/ML practitioners.

Finally, the leadership of these four functions is crucial. BI, AI/ML Ecosystem, AI/ML Product, and AMI functions need to establish a collaborative environment. Also, the leadership of the data team and C-level leadership need to establish a foundational understanding of the AI/ML discipline. Introducing a dedicated C-level leadership role for AI/ML and data might be beneficial. This organizational strategy might need to evolve as an organization increases its focus on AMI. Nevertheless, all the corresponding functions in an organization will need to be defined to increase the chance of a successful outcome. Additionally, the Engineering -and Data team needs to align on the resources, deliverables, roadmap, etc., in order to support business goals. The C-level leadership should review this alignment (see figure~\ref{fig:AI_ML_organigram}).

\section{Strategy and planning for AI/ML Innovation (AMI)}
\label{section:strategy_planning_ai_ml_innovation}

The C-level leadership of an organization needs to find a balance between exploitation of existing capabilities related to the core business, i.e., products and services, and exploration of future capabilities related to new business opportunities \cite{duncan1976ambidextrous,march1991exploration}. The co-existence of exploitation and exploration is a cornerstone for organizational ambidexterity \cite{tushman1996ambidextrous}. Exploitation is accomplished by incremental -and sustaining innovation and innovation as a process. The goal of this exploitation is to illustrate operational efficiency. Alternatively, exploration results in radical -and disruptive innovation and innovation as something new and innovation as invention \cite{birkinshaw2004building}. The goal of the exploration is to enter new markets or disrupt existing markets. There are two main approaches to achieve an ambidextrous organization: (1) \emph{structural ambidexterity}: exploitation and exploration are separated in the organigram and (2) \emph{contextual ambidexterity}: each employee balances their time between exploitation and exploration. Since AMI demands high levels of expertise, structural ambidexterity is often preferred (see Figure~\ref{fig:AI_ML_organigram}).

\subparagraph{Competitive advantage with Porter's five forces} A business strategy focuses on achieving a competitive advantage. Porter's five competitive forces define the attractiveness of an industry: (1) the threat of new entrants, (2) the bargaining power of buyers, (3) the bargaining of suppliers, (4) the threat of substitute products and services, and (5) the rivalry among existing competitors \cite{strategy1979michael}. In the last decades, the business world has transformed due to globalization; consequently, the hyper-competitive business environment of the 21st-century demands for reviewing Porter's five forces \cite{arent2016five,isabelle2020porter}. More recently, additional forces proposed in the literature: the competitor's level of innovativeness, exposure to globalization, the threat of digitalization, and industry exposure to de/regulation activities, data-driven -and IT-enabled solutions, radical innovation at speed and scale, hyper-effective use if assets, climate resiliency and environmental stewardship, social engagement, etc. Therefore, competitive advantage is embedded into a framework for disruptive AMI.

\subparagraph{Funding} In order to support innovation sustainably, it is vital to secure funding to support the gradual maturity of innovative tracks of work. This funding can come from internal -or external entities. The process of maturing innovative work can be long and labor-intensive. Therefore, funding is crucial for maturing early-stage developments that show promising performance. This funding can come from the exploitation of existing products and services. Therefore, it is essential to forecast the profits and secure competitiveness of the existing products \cite{tushman1996ambidextrous}. 

\begin{figure}[!ht]
\centering
\includegraphics[scale=0.8]{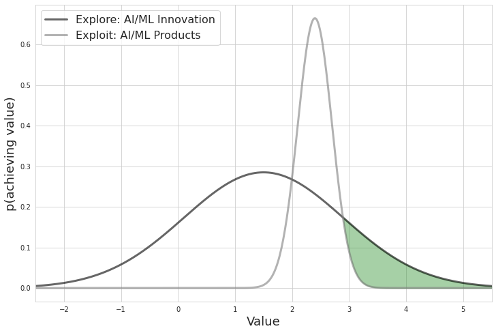}
\caption{The different between an established AI/ML Product and a novel AI/ML Innovation (AMI) initiative based upon its potential value and  the probability of achieving its potential value.}
\label{fig:AI_ML_value}
\end{figure}

As for other technologies, the exact characteristics of exploiting existing AI/ML Products and exploring novel AMI initiatives occurs. However, the AI/ML field has a subtle difference: AMI is a given. Moreover, since AI/ML is a rapidly changing research field, new AI/ML Products or significant AI/ML products extensions are expected. This evolution enables new computational frameworks and hardware that galvanizes the speed of novel algorithm development. Consequently, changes in DataOps and MLOps solutions are happening. The leadership of an organization should manage the disruption related to innovation in the organization. Today, the leadership tries to understand and manage the \emph{great resignation} of talent. Illustratively, high levels of innovation are reported as the third highest driver for attrition \cite{sull2022innovation}.

\subparagraph{Uncertainty} Unconsciously, we might be following the black swan theory \cite{kim2020}. Since the outcome of the innovation is uncertain, it would be safe to invest in experimentation tools for evaluating the performance of AMI initiatives. Subject matter experts and AI/ML experts might have a strong intuition for the likelihood of success for a novel algorithm. Nevertheless, the higher the levels of innovation deal with a higher risk but higher potential reward (see the green area in Figure~\ref{fig:AI_ML_value}). Therefore, guiding risks, redefining success throughout the AMI journey, building generic building blocks across initiatives, defining exit strategies, etc., are essential for maturing AMI within an organization. Patience for the few successes is an essential mindset for AI/ML practitioners -and leaders and the C-level leadership.

Defining in which new AMI initiatives to invest in is rather complex \cite{gailly2018navigating}. There might be multiple solutions, various dependencies, implications for sustainable growth, etc., that decision-makers need to consider. Generally, it is essential to merge a shared strategic vision related to the value chain, supply chain, and ecosystem with disruptive external trends in the market, developing a balanced portfolio and delivering added value to customers. Since innovation is the process of generating economic value after the invention, in the following sections, we will review the AMI journey and a tool for defining its maturity: technology readiness levels \cite{price200612}. 

\section{AI/ML Innovation (AMI) Journey}

\begin{figure}[!ht]
\centering
\includegraphics[scale=0.8]{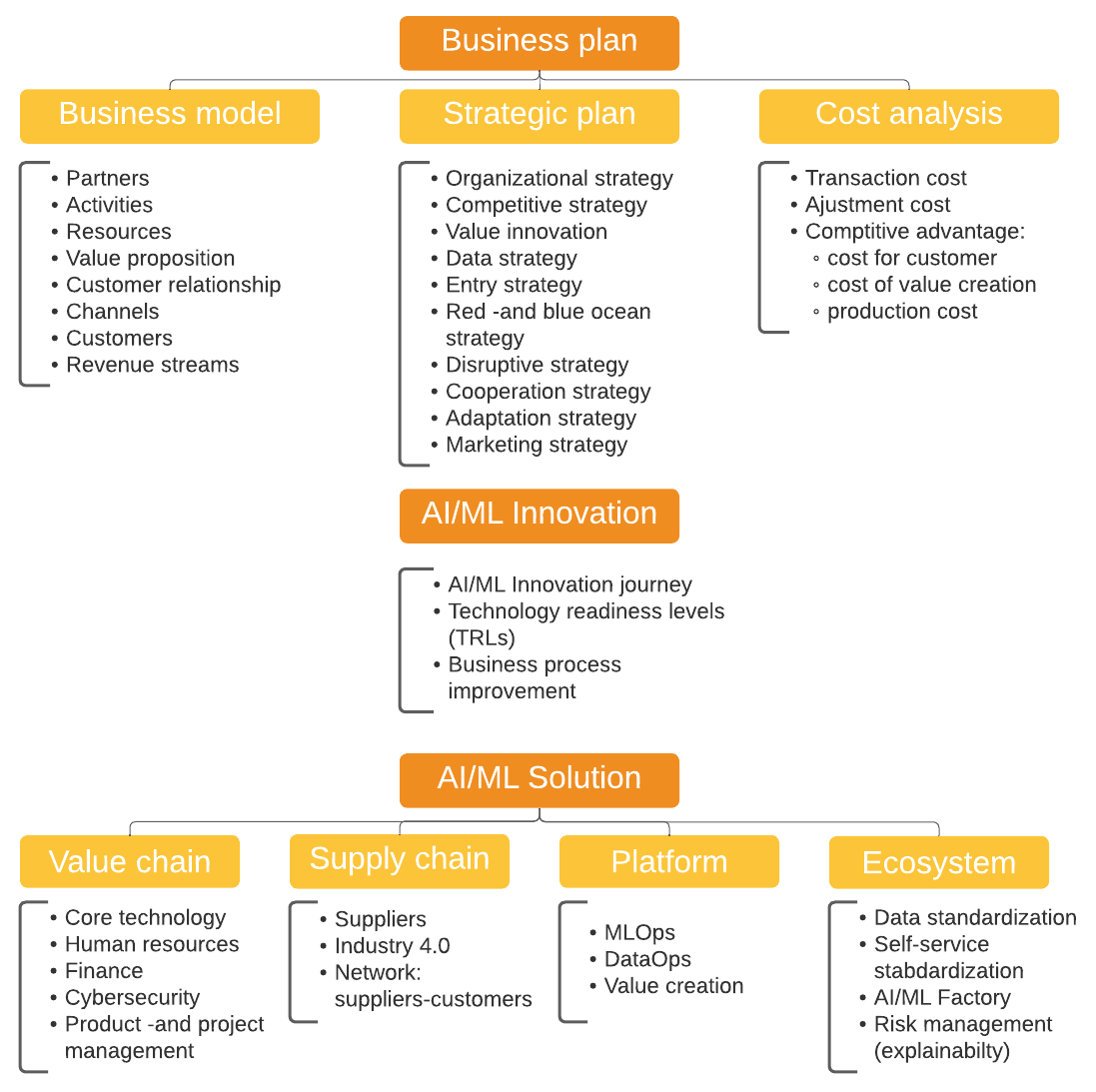}
\caption{Overview of the AI/ML Innovation (AMI) framework.}
\label{fig:AI_ML_framework}
\end{figure}

The journey for an AMI initiative can help understand the transition from ideation to AI/ML Product. Defining a format and a forum for bottom-up and cross-functional intent reviews is a cornerstone to define which new AI/ML Product to build. Establishing a regular cadence of participants across the organization might be an excellent procedure to stimulate novel AI/ML Product development proposals. These proposals should align with the business plan of the organization (see figure~\ref{fig:AI_ML_framework}).

\begin{figure}[!ht]
\centering
\includegraphics[scale=0.56]{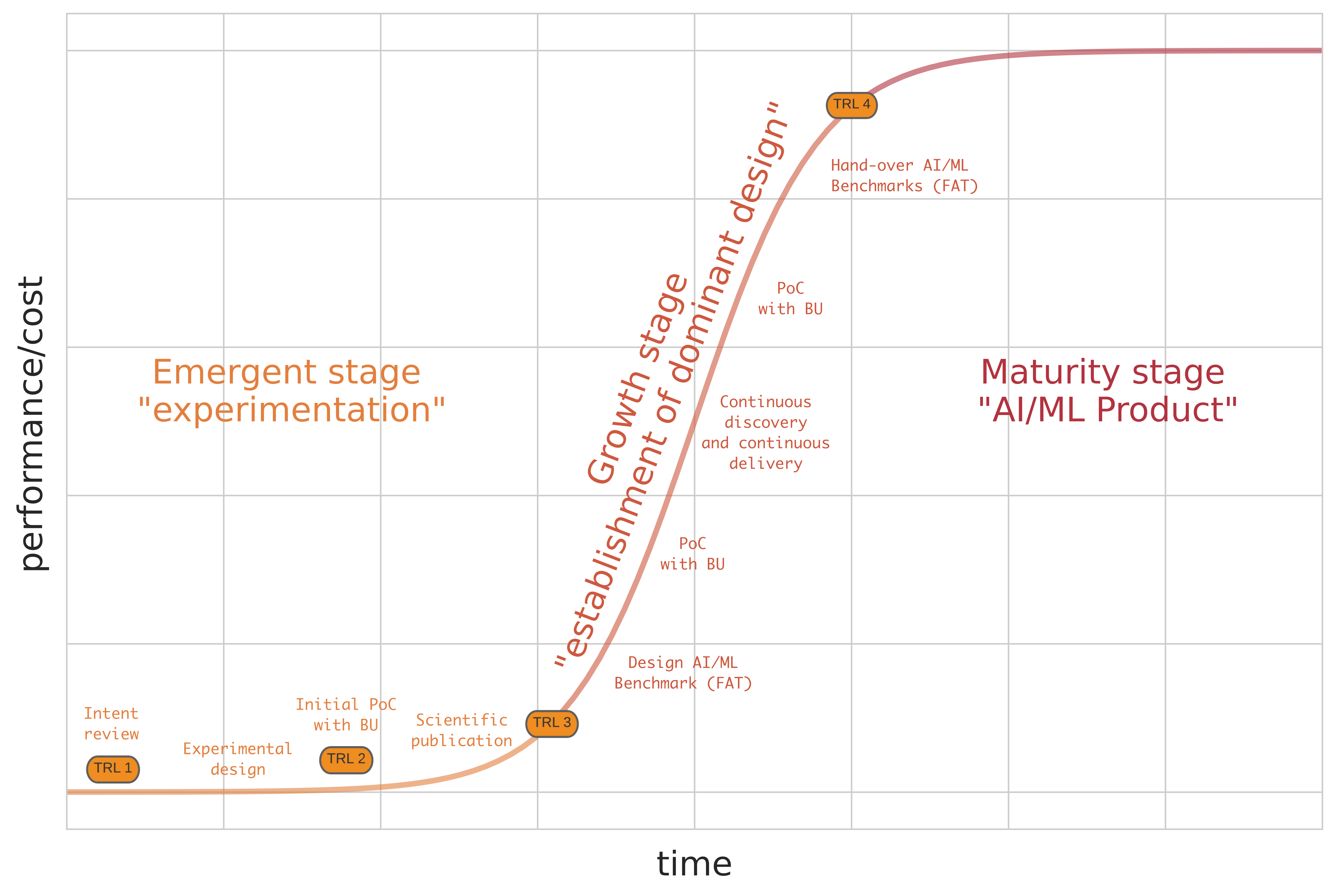}
\caption{The technological evolution of an AI/ML Innovation (AMI) initiative.}
\label{fig:AI_ML_tech_evolution}
\end{figure}

The technology evolution is often visualized as an S-curve \cite{sahal1981patterns,brown1992managing}. The S-curve framework illustrates the rate of progress in performance during technology development. There are three main stages during technology development (see figure~\ref{fig:AI_ML_tech_evolution}): (1) the emergent stage represents the experimentation during early-stage development of new technology, (2) the growth stage represents the establishment of a dominant design, and (3) the maturity stage represents the operational efficiency of the new technology. The S-curve applied as a forecasting tool should be critically assessed \cite{foster1986innovation,christensen1992exploring}. During the growth stage, the new technology is better understood and easier to use, its rate of progress increases. During the emergent and maturity stage, the rate of progress is relatively low and costly.  

A tool to measure the maturity of technology is called technology readiness levels (TRLs; see Figure~\ref{fig:AI_ML_tech_evolution}). The TRLs defined by NASA \cite{Tzinis2021} and European Union \cite{wiki2021} contain nine levels. In the following section, the first four levels will be redefined to capture AI/ML Product development progress. Also, these TRLs could introduce standards for AI/ML Products. i.e., coding -and data standards, documentation availability, business communications, etc. As the technology improves in maturity, standardization of the AMI journey could help further maturity. During the emergent stage, this standardization could delay experimentation.

\subsection{Technology readiness levels (TRLs)}

This section will review the four technology readiness levels (TRLs) during AI/ML Product development. This AMI journey will result in a technology transfer to the AI/ML Product team. The maturity stage (TRL5 until TRL 9) will be out-of-scope of this manuscript. The maturity stage will be performed by the data team's AI/ML Product function.

\subparagraph{Review current solutions} The first TRL is called \emph{review current solutions}. This level identifies a set of existing algorithms that would be good candidates to predict an insight or perform a computational task necessary for the AI/ML Product. The selected algorithms can be an older baseline solution previously used for the task or a novel published solution. We need to reproduce the most relevant results of these algorithms. It is important to note that we need to verify if the invention, i.e., the intellectual property (IP), software license, etc., is compliant with the IP policy of the organization.

\subparagraph{Define experimental design for comparing innovative solutions} The second TRL is called \emph{define experimental design for comparing innovative solutions}. This level uses the experimental design for evaluating the performance of baseline, published, and novel solutions for a computational task performed by an algorithm. This performance evaluation is frequently performed for publicly available data; these data resources are used in the corresponding literature. The implementation of the experimental design and the corresponding critical assessment of the performance of the algorithms will close the second TRL. A scientific publication can be submitted if the proposed solutions are novel enough.

\subparagraph{Use case translation} The third level is called \emph{use case translation}. This level uses the algorithms developed during TRL 2. Each algorithm is benchmarked against their real-world or proprietary data. A critical assessment of these algorithms against this data will provide insight into the translated performance by executing the experimental design. This experimental design will evolve into an AI/ML benchmark, including validating and verifying the AI/ML solution with customers and end-users. Finally, a DataOps -and MLOps architecture is defined to enable reuse of the AI/ML Benchmark for future deployment and use cases. TRL 3 is closed by implementing the AI/ML Product prototype: the DataOps and MLOps capabilities, the AI/ML algorithms, and the AI/ML Benchmark.

\subparagraph{Prototype validation} The fourth level is called \emph{prototype validation}. Before reaching TRL 4, the AI/ML Product prototype will be applied for several use cases. The goal is to reach a dominant design for the AI/ML Product. TRL 4 is reached when the AI/ML Product has shown consistent performance across many use cases. Finally, the DataOps and MLOps architecture requirements are fully defined.

\subsection{Commercial ideas}
\label{section:commercial_ideas}

Generating and evaluating commercial ideas is an important responsibility of C-level leadership. Therefore, scanning and learning new technologies in the market is key for making sound decisions. Consequently, C-level executives need to make \emph{"build versus buy"} decisions in order to achieve business goals \cite{fowler2004build}. The leadership must estimate the investment for building an AI/ML solution related to a business use case, i.e., the preparation of the underlying data for the AI/ML solution, the evaluation of the optimal parameters for AI/ML Product for this use case, hosting the insights of the AI/ML solution to the business, maintaining the quality of the insights of the AI/ML solution over time, etc. Additionally, there will be a need to do some degree of buying. More and more commercial solutions for DataOps and MLOps dependencies of an AI/ML solution. Furthermore, buying an AI/ML solution can lead to a merge and acquisition of another firm, define an alliance with another firm, invest in a spin-out, etc. AI/ML or data is not part of the core technology portfolio in some organizations. Therefore, a \emph{data strategy} should identify crucial data assets and the corresponding commercial ideas for applying AI/ML in the organization \cite{constantiou2015new}. Additionally, this data strategy prioritizes efforts for organizing, governing, analyzing, and deploying an organization’s information assets \cite{dallemule2017s,nyc2021aistrategy}. All C-suite executives play an instrumental role in making progress in the strategy related to data management.

The C-level leadership should attempt to orchestrate a holistic product portfolio vision with adequate attention for AI/ML and data. Additionally, AI/ML and data have made massive progress in importance for many organizations \cite{Bhattacharya2021}. However, despite the progress, the funding for AI/ML and data in an organization might be challenging. Therefore, the C-level leadership and the AI/ML leadership should define processes for generating new ideas and identifying commercial ideas that need AI/ML and data solutions.

One area for leveraging AI/ML and data is right at the core of the business strategy and the corresponding corporate scaling \cite{kiron2019strategy,Chuprina2020,Mulligan2021,Vesuvala2021}. Any form of expansion, i.e., geographic, vertical, horizontal, etc., could be better informed by data-driven decisions. Researching demand for new efforts of expansion could inform decision-makers with additional information \cite{Chuprina2020}. More generally, AI/ML and data can complement the current business practices with computational methods for advising organizations about mergers and acquisitions, new products and services, customer targeting, finance, etc. AI/ML and data could also result in better business development processes \cite{daugherty2018human}. 

\subparagraph{Environmental, social, and governance (ESG) factors} More recently, many investors evaluate organizations based upon additional criteria related to sustainability \cite{van2016esg}. A general way to encapsulate the sustainability of an organization is by reviewing environmental, social, and governance (ESG) factors. An environmental factor is related to climate change, resource depletion, pollution, etc. The social factor represents the organization's position related to modern slavery, child labor, diversity and inclusion, etc. In an organization, a governance factor describes the prevention of excessive executive pay, a fair tax strategy, the ethical considerations related to avoiding bribery and corruption, etc. There are two domains AI/ML can contribute to ESG: (1) the organization could define measurable metrics and use analytics in order to monitor the progress on ESG \cite{chen2020integrated} and (2) the organization could measure the ESG footprint related to AI/ML \cite{saetra2021framework}. Furthermore, ESG starts to be critical criteria for evaluating novel technologies \cite{liu2021blockchain,lai2021blockchain}. The innovation management in an organization can be improved through AI \cite{haefner2021artificial}. During the innovation process, AI can help the generation of ideas and the development of ideas (see section~\ref{section:introduction}). More generally, the automation of business processes can be accomplished by process mining and robotic process automation (RPA) methodologies \cite{van2012process,leno2021robotic}. This level of automation is possible by collecting information in a digitized organization. Additionally, generative design has shown examples of effective product customization \cite{laura2021product}.

\subsection{Entry strategies}
\label{section:entry_strategy}

An important aspect of AI/ML Product development relates to the definition of an entry strategy. The AMI might want to concentrate on collaborations with internal -or external incumbents related to the supply chain, value chain, and ecosystem strategy. During the early stages of the AI/ML Product development, it can be beneficial to collaborate across the organization to enable early adoption by future users, adaptation into an ecosystem platform, and avoid duplication. Entry strategies should include funding-exit decisions \cite{basole2021visualizing}.

\subparagraph{Red -and blue ocean strategy} There are two distinct entry strategies for a novel AI/ML Product: (1) red ocean strategy: an existing market space or (2) blue ocean strategy: a new market space \cite{mauborgne2006blue}. In organizations with a more mature AI/ML function, new AI/ML Products proposed by members of the organization (the Board for Innovation; see section~\ref{section:InternalBusiness}) might be variants, improvements, or complements of existing AI/ML Products. Therefore, the AI/ML teams have strong expertise in AI/ML for a specific market and the future application of the AI/ML Products. Consequently, they will often define a red ocean strategy. Alternatively, in organizations with a less mature AI/ML function, novel AI/ML Products development could be more greenfield and should be approached differently. More effort on user research, customer acquisition, organizing training and learning of product team members, targeted hiring, etc., are crucial for future success. Therefore, these organizations will often focus on a blue ocean strategy \cite{agnihotri2016extending}.

\subparagraph{Disruptive -and cooperation strategy} In some cases, a novel AI/ML Product may need a disruptive strategy; this strategy often emphasizes complementary capabilities to existing AI/ML products. Regularly, a new AI/ML Product can outperform existing AI/ML Products; therefore, the emergent AI/ML Product disrupts the existing AI/ML Product market. Therefore, appropriate actions should be taken to accomplish a seamless transition. A framework for entry strategies proposes two approaches \cite{gans2003product}: (1) the organization should operate in the product market with manufacturing, marketing, and distributing innovative products or (2) the organization defines a \emph{cooperation strategy} by defining a market of ideas with commercializing an innovative product through licensing its rights. This framework is mainly instrumental for positioning innovation in new technologies. Furthermore, an organization can collaborate with universities to develop innovations for novel AI/ML Products.

\subparagraph{Transaction cost} In recent decades, much research has focused on defining the transaction cost, also called the cost of doing business \cite{coase1937nature,williamson1979transaction}. The estimation of the transaction cost is still one of the most active research areas in modern economic science. Potential drivers of transaction cost are, i.e., cost of administrative barriers overcoming, information search costs, negotiation costs, costs of measurement, costs of specification and property rights protection, costs of opportunism, and lobbying costs \cite{cuypers2021transaction}. Defining software licensing costs, developing generic AI/ML Products, prioritizing less complex AI/ML Product development, etc., can lower transaction costs. The transaction cost related to switching MLOps and DataOps dependent technologies could be essential to consider. Additionally to the transaction cost, the production cost related to deploying future AI/ML Products can be informative for defining the entry strategy.

\section{Strategic adaptation and renewal}
\label{section:strategic_adaptation_renewal}

Generally, the adaptation of new technology represents the trade-off between the sustainability of existing technologies and the disruption originated from emergent technologies \cite{christensen2013innovator}. Therefore, the evaluation of the impact of emerging technology on an organization's business model and assets are on the radar of many technology leaders. Additionally, many research reports on the adjustment costs related to the adaptation of emergent technologies \cite{lucas1967adjustment,basu1987adjustment,meghir1996job}. Estimating the adjustment cost for adapting AI/ML helps prioritize and identify future AMI initiatives. Potentially, AI/ML adaptation impacts an organization heterogeneously; since it can represent multiple scenarios of the Innovator's Dilemma \cite{christensen2018disruptive}. Defining an adaptation strategy for an organization will be instrumental for the future adoption of AI/ML Products and services. This strategy should guarantee that different business units, teams, functions, etc., of an organization, are prepared to embed AI/ML into their business practices -and processes. Furthermore, this adaptation strategy should define an IP strategy assessing the product ownership and the intellectual -and data property \cite{pasquinelli2adapt}. 

\subparagraph{Value innovation} Often, the disruption related to emergent technologies focuses on the response of an incumbent organization for the competitors providing a better substitute in the market. In AMI, this disruption is the default; therefore, the response to disruptive innovation is a phenomenon that all organization members should embrace. Therefore, it is essential to stimulate an innovative culture in an organization \cite{martins2003building,dobni2008measuring}. In management literature, the disruption by emergent technologies is categorized by two types of threats \cite{adner2021disruption}: (1) threats among the existing set of competitors, i.e., internal dysfunction and competitive rivalry and (2) threats outside of the existing competitors, i.e., substitute products and complementary products. Additionally, the conventional strategy and strategic frameworks for disruptive innovation focus on achieving competitive advantage \cite{strategy1979michael,christensen2018disruptive}. Alternatively, the framework based upon \emph{value innovation} promises higher growth by focusing on superior value for the customers and their offerings in the market \cite{kim1997value,kim1999strategy}. However, focusing only on outperforming competitors could result in organizations neglecting their customers' most essential needs for products and services.

This manuscript suggests that defining a business strategy for disruptive innovation based upon competitive advantage has limitations for AI/ML. A framework for AI/ML disruptive innovation is composed of components vital for the digital transformation of the supply chain, value chain, and ecosystem. In the following sections, we discuss the AI/ML disruptive innovation: (1) internal business practice (value chain) and (2) external third parties and customers (supply chain, platform, and ecosystem).

\subsection{Internal business practice}
\label{section:InternalBusiness}

\subparagraph{Board on innovation} The C-level leadership might want to build a \emph{Board for Innovation} within the firm \cite{chen2016ownership,hill2017board,balsmeier2017independent,board2019ai}. Too often, innovation is initiated by demand for help. Potentially, this can lead to a lack of vision and strategy even if the AI/ML leaders proposed a vision and a strategy. Its execution can be interrupted by a demand to overcome critical liabilities and a lack of competencies within an organization. Additionally, some research suggests that an independent board of directors can lead to more effective innovation \cite{balsmeier2017independent}. Therefore, a dedicated Board for Innovation can provide a centralized entity in an organization focused on managing the tension to stimulate thinking, ideas, and innovation. If there is no Board of Innovation, the C-level leadership should organize and identify a replacement. One of the Board of Innovation goals is to define an innovation portfolio. Defining this portfolio could be accomplished by organizing an innovation tournament \cite{terwiesch2009innovation}. This tournament intents to initiate a broad set of potential innovation initiatives. The Board of Innovation should define the criteria for selecting the most promising initiatives for an innovation portfolio \cite{wooten2017idea}.

\subparagraph{Employee network} A crucial part of an organization is the network of employees \cite{yang2021cascaded}. Enabling employees to search for other employees in the organization enables potential collaborations and a better understanding of reporting structures. In addition, the creative disciplines often are fragmented across different teams; therefore, networks are a great tool to map out practitioners of these disciplines.

\subparagraph{Activity system map} An activity system map \cite{Porter1996,Maital2016} of your organization is a visualization tool that connects the value proposition of an organization with the activities within your organization. Consequently, activity system maps allow analyzing the competitive advantage of an organization \cite{richardson2005business}. Furthermore, an activity system map provides the focus areas for building an ecosystem \cite{schneckenberg2021theorizing}.

\subparagraph{Competitive strategy cost} Competitive advantage can impact two types of cost: (1) the cost for the customer (cost optimization) and (2) the cost of the values created by an organization (cost reduction). In the context of the internal business practice for AMI, it is crucial to concentrate on areas that can improve cost reduction. Many organizations have core functions, i.e., human resources, project management, product management, software engineering, finance, marketing, sales, etc. Each of these functions could profit from AMI initiatives. A Board of Innovation, or its replacement, could guide the process of prioritization, funding, and alignment with the organization's leadership. It is essential to synchronize these innovation efforts with the business model, the strategic plan, and cost analysis of the organization \cite{kiron2019strategy}.

\subsection{External 3rd parties and customers}
\label{section:ExternalCustomers}

\subparagraph{Marketing strategy} External third parties and customers are a continuation of the internal business practice. The framework for disruptive AMI would also need to focus on the pricing optimization of the AI/ML Product for the customer. Furthermore, defining the next-generation features of a product should incorporate the role of a new AI/ML Product. Now, the organization can implement the marketing of a new AI/ML Product. Therefore, a \emph{marketing strategy} is fundamental for building a brand. Also, searching for prospects and defining customer segments of the product portfolio is essential \cite{ali2021marketing}. Additionally, digital marketing can support the marketing strategy \cite{conick2017past,mariani2021ai,van2021artificial}. Innovation and its related marketing are also crucial in the race for global AI/ML talent. Since talent in AI/ML is sparse and often clustered, it is crucial to illustrate the impact of AI/ML in a company.

A shared vocabulary for AI/ML helps an organization develop a business plan. More recent research introduced the term AI Factory \cite{jia2021artificial} as a catalog of AI/ML Products. However, reviewing customer-facing AI/ML initiatives is essential to provide a precise vocabulary for better communication with the supply chain, platform, and ecosystem actors.

Similarly, with the internal business practice, the prioritization by a Board of Innovation would centralize the AMI initiatives related to external third parties and customers. In addition, this would guarantee that expert knowledge is incorporated by the decision-making \cite{board2019ai}.

\subparagraph{Customer adoption} An important aspect for new AI/ML Product development is customer adoption \cite{cubric2020drivers}. There are two main drivers for customer adoption: (1) the prediction provided by the AI/ML algorithm to the customer with satisfactory performance (insight in figure~\ref{fig:AI_ML_product_development}) and (2) the thrust that the customer has in the predictive capabilities of the AI/ML algorithm. The AI/ML Product -and AI/ML Ecosystem functions implementing the AI/ML solution should closely monitor insight performance, and the business intelligence teams will analyze cannibalization, customer retention, etc. AI/ML safety and explainability can differ depending on the marketplace, i.e., health care, multimedia, aviation, etc., and the use case. 

In order to overcome barriers for AI/ML Product adoption, the user of the insights provided by the AI/ML Product needs sufficient training and education in order to reduce the fear of job loss and dependence on non-humans \cite{cubric2020drivers}. Furthermore, early customer involvement can inform AI/ML practitioners in an organization about user-centric goals. During the early stages of the AMI journey (TRL $\leq$ 2), customer feedback can provide instrumental input to learn and fail fast. Incorporating regular customer feedback by the Agile methodology will construct the correct AI/ML solution.

\begin{center}
\begin{table}
\begin{tabular}{p{4cm}|p{9cm}}
Strategy type & Description \\
\hline
Organization strategy & The long-term strategic actions implemented by an organization, i.e., mission -and vision statements, business -and functional objectives, resource management, scope, and core competence of the organization (see section~\ref{section:organizational_strategy}). \\
\hline
Competitive strategy & The traditional business strategy aims the achieve long-term sustainable competitive advantage (see section~\ref{section:strategy_planning_ai_ml_innovation}). \\
\hline
Value innovation & This strategy focuses on value creation for customers and the market; its purpose is achieving higher sustainable growth compared to competitive strategy (see section~\ref{section:strategic_adaptation_renewal}). \\
\hline
Data strategy & The data strategy defines the data management of an organization, i.e., data assets and the corresponding commercial ideas for applying AI/ML, the priorities for organizing, governing, analyzing, and deploying an organization’s information assets (see section~\ref{section:commercial_ideas}). \\
\hline
Entry strategy & The entry strategy defines the optimal path for new product development adoption in an organization. Consequently, this entry strategy should also cover novel AI/ML Product developments (see section~\ref{section:entry_strategy}). \\
\hline
Red -and blue ocean strategy & These entry strategies are distinct based upon the market for novel product development; a red ocean strategy focuses on an existing market space, and a blue ocean strategy focuses on a new market space (see section~\ref{section:entry_strategy}). \\
\hline
Disruptive strategy & The strategy focused on the replacement of established technology by emergent technologies (see section~\ref{section:entry_strategy}). \\
\hline
Cooperation strategy & This entry strategy focuses on the collaboration between organizations; often, it is called the market of ideas. As a result of this, the incumbent supports the idea for novel product development (see section~\ref{section:entry_strategy}). \\
\hline
Adaptation strategy & This strategy focuses on the adaptation of novel technologies and products in an organization (see section~\ref{section:strategic_adaptation_renewal}). \\
\hline
Marketing strategy & This strategy focuses on the marketing of the product portfolio and the brand of an organization (see section~\ref{section:ExternalCustomers}). \\
\end{tabular}
\caption{\label{tab:strategic_plan}The strategic plan of an organization.}
\end{table}
\end{center}

\subsection{Data privacy and AI regulation}

Data privacy and AI regulation will gain importance in the next decade. As AI/ML is gaining importance, its data privacy and AI regulation will evolve. Unfortunately, regulation is often behind the technology. Currently, this regulation shows significant geographic differences. Therefore, it is crucial to define a scalable compliance and cybersecurity process related to the ingested data in the ecosystem. 

\subparagraph{Data privacy} Data privacy is an essential topic for C-suite leadership. Data privacy regulates how data is collected, stored, shared, and deleted \cite{chauhan20212021}. Therefore, a data privacy policy that users, who share their data, can agree upon is an important cornerstone. Additionally, C-suite leadership should define the core principles and the code of ethics for using technology in an organization. Furthermore, implementing privacy-by-design for better-standardized solution architectures improves automation during software development, and security \cite{khashooei2021architecting}.

\subparagraph{Section 230 of the CDA} In 1996, section 230 of the Communications Decency Act (CDA) attempts to protect freedom of expression and innovation of the internet. It states: "No provider or user of an interactive computer service shall be treated as the publisher or speaker of any information provided by another information content provider\footnote{47 U.S.C. section 230: \url{https://www.law.cornell.edu/uscode/text/47/230}}". Section 230 protects internet service providers (ISPs) and a range of cloud services providers (CSPs) from being responsible for the content they provide \cite{kosseff2019twenty}. In recent years, misinformation, disinformation, harassment, hate speech, etc., have put section 230 of the CDA under pressure. Publishers of video -and audio content and social media platforms receive more requests to moderate their content. Some of these publishers have attempted to filter content. The lack of an independent regulatory framework resulted in public debate and demand by these private companies for more guidance for this moderating role for filtering content.

Furthermore, internet users have become increasingly aware of the importance of their data and their personally identifiable information (PII). The recent pandemic has also increased the use of social media and video communication platforms to stay connected during lockdowns. In addition, many organizations allow to work from home, more frequently used eCommerce, new contact tracing services to manage the spread of COVID-19, etc., have resulted in the increased focus on data privacy and security.

\subparagraph{GDPR} In 2018, the European Union introduced the General Data Protection Regulation (GDPR). The GDPR balances personal data protection and data processing needed for innovation \cite{syroid2021personal}. In GDPR article 4, personal data is defined as: "any information relating to an identified or identifiable natural person (‘data subject’); a natural person can be identified, directly or indirectly, in particular by reference to an identifier such as a name, an identification number, location data, an online identifier or to one or more factors specific to the physical, physiological, genetic, mental, economic, cultural or social identity of that natural person." The GDPR is an integral part of human rights and privacy laws in the EU and is applicable for any organization that provides goods or services in the EU. Every resident in the EU has the right to access, the right to rectification, restrict processing, portability, and the right to object concerning their personal information. The GDPR impacts the way organizations need to perform data governance and system design. An initial step in data governance is to ensure all business activities are mapped with PII data: corresponding data flows, data owners, compliance reviewers, and stakeholders. 

Traditionally in the US, organizations have more autonomy in handling PII data. Potentially, there are a set of federal laws regulating different types of data, i.e., Health Insurance Portability and Accountability Act (HIPAA), Fair Credit Reporting Act (FCRA), Family Educational Rights and Privacy Act (FERPA), Gramm-Leach-Bliley Act (GLBA), Electronic Communications Privacy Act (ECPA), Children's Online Privacy Protection Rule (COPPA), Video Privacy Protection Act (VPPA), and Federal Trade Commission Act (FTC Act). Additionally, the antitrust law is gaining attention for the digital economy \cite{douglas2021digital}. Significantly, the dominance of Google, Apple, Facebook (Meta), Amazon, and Microsoft (GAFAM) stock raises concern from governments about customer protection from predatory business practices and ensuring fair competition. Furthermore, different states in the US started to regulate data privacy, i.e., California Consumer Privacy Act (CCPA) and California Privacy Rights Act (CPRA), Virginia's Consumer Data Protection Act (CDPA), Colorado Privacy Act, etc. Data privacy laws are a collection of federal -and state laws. Therefore, an organization must seek appropriate legal advice for the implementation of corresponding security solutions \cite{ibmsec2021}.

\subparagraph{AI regulation} In 2017, China introduced a new strategy: \emph{'New Generation Artificial Intelligence Development Plan'} after Lee Sedol lost an exhibition Go game against Google's AI algorithm AlphaGo by four games to one watched by over 280 million \cite{roberts2021chinese}. In 2018, the European Commission published an AI strategy: \emph{'lead the way in developing and using AI for good and for all'}. In 2019, the US AI strategy called: \emph{'accelerate the nation's leadership in AI'}. Many other countries published AI strategies in recent years. More recently, many regulators and authorities have started to realize that there is 'no such thing as a free lunch' \cite{friedman1975there}. The benefits of AI/ML can be significant; nevertheless, the legal and ethical implications of using AI safely and trustworthy are equally important. Consequently, the focus on AI strategy has shifted into AI regulation \cite{smuha2021race}.

Often, there are four modalities for introducing regulation related to new technology \cite{lessig1991horse}: (1) the law, (2) the society, (3) the market, (4) the architecture or design of technological applications. Authorities need to balance the protective role and the enabling role of regulation. Furthermore, these modalities and the balance between the protective -and enabling role of regulation interplay requires a holistic review. 

In 2021, the European Commission (EC) developed a proposal for an AI regulation called the Artificial Intelligence Act. The protective role of this legislation addresses manipulative conduct, indiscriminate surveillance, and large-scale social scoring. In the meantime, China has developed a large-scale credit score system including AI/ML technology \cite{hansen2021universalizing}. Many governments are critically assessing such large-scale scoring AI/ML scoring systems. Not only if it is legal, but defining trustworthy AI or responsible AI. How can we define ethics and robustness? First, an AI regulation needs to answer demands for more transparency. Therefore, explainability and interpretability of the AI systems, standardization for AI//ML system vendors, privacy and non-discrimination metrics, etc., are essential to regulate the social impact of an AI system. Also, in the US, there is an Algorithm Accountability Act in 2019 and the Algorithmic Justice and Online Platform Transparency Act in 2021 that would require organizations to make impact statements. This impact statement mandates that organizations report details about their AI systems related to their risks and how to mitigate these risks. The impact statement must be completed before the AI system is deployed, and the statement might need to be made public and available for regulators and audits. This regulation in the US is not yet fully adopted.

\section{Digital transformation}
\label{section:digital_transformation}

\begin{figure}[!ht]
\centering
\includegraphics[scale=0.56]{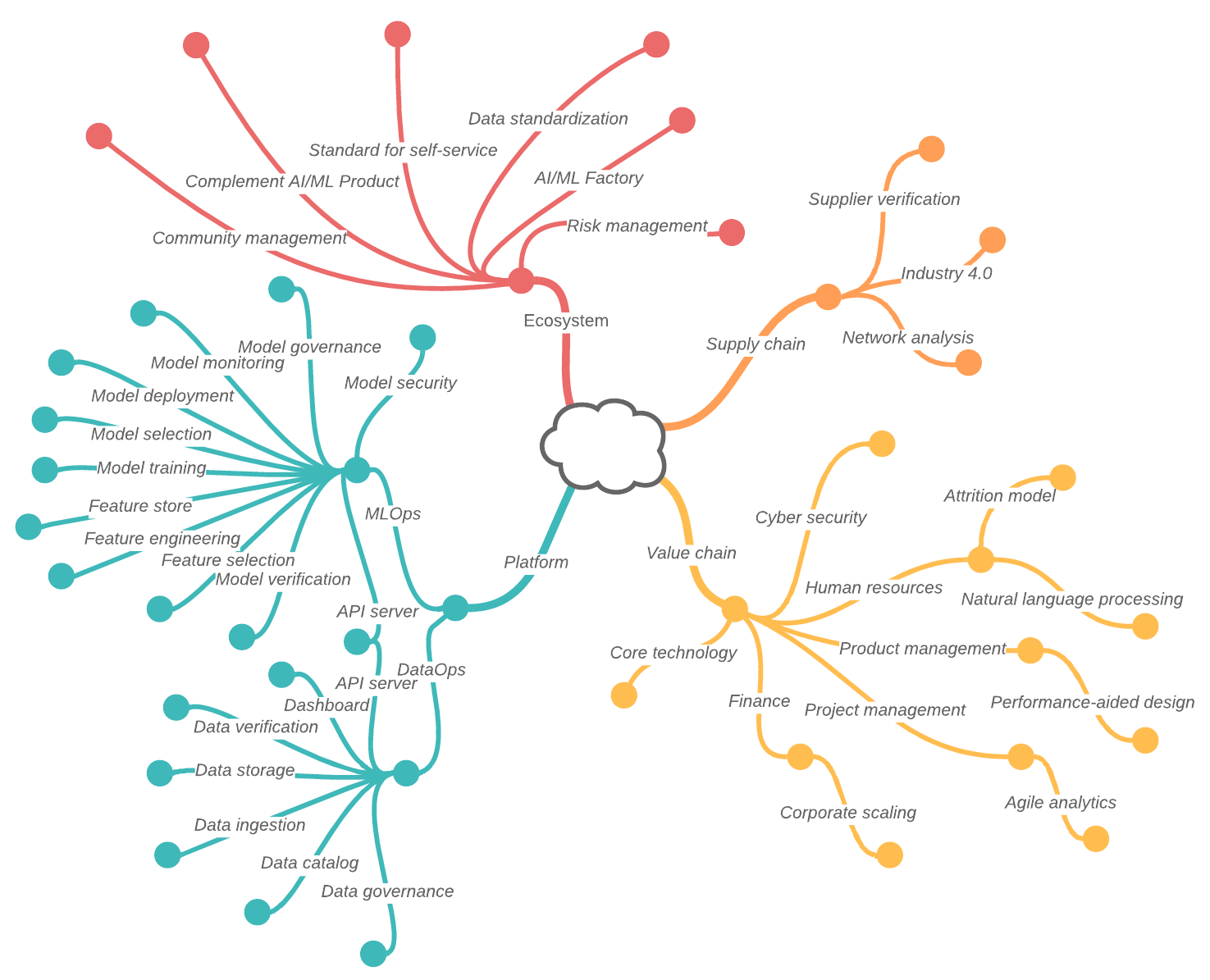}
\caption{A mind map of the disruption on the strategic adaptation and renewal for AI/ML Innovation (AMI) in an organization.}
\label{fig:AI_ML_mindmap_disruption}
\end{figure}

\subparagraph{Ecosystem} Traditionally, supply chains and value chains define an organization's business activities. However, these business goals are redefined to date to accomplish digital transformation. Additionally, one of the essential concepts of digital transformation is an ecosystem \cite{moore1993predators,sumbaly2013big}. An ecosystem is a term that originates from ecology \cite{tansley1935use}. The interaction of species with their biotope has been studied extensively by ecologists; the climate change crisis has reignited interest in ecological ecosystems \cite{etde_6041139}. There are four main types of interactions in these ecological communities \cite{Michigan2008}: (1) mutualism: 2 species benefit, (2) commensalism: one species benefits and the impact for the other species is neutral, (3) competition: each species affected negatively, and (4) predation: one species benefits and the other species is affected negatively. Since the increased complexity of products and the sustainable growth of organizations effectively incorporating ecosystems in their business strategy, much attention in the management literature goes to business ecosystems \cite{adner2010value,hamel2018end}. A business ecosystem seeks sustainable growth by mutualism.

\subparagraph{Multiple S-curves} During a digital transformation effort, leadership needs to decide on technologies for different components and architectures; consequently, each of these technologies represents an S-curve. Therefore, an AL/ML solution will be a composition of multiple S-curves \cite{adner2016innovation}. Additionally, a solution for digital transformation will need to adopt new emergent technologies over time. The expected disruption from emergent AI/ML Products for supply chains, value chains, and ecosystems are reviewed in the following sections.

\subsection{Value chain and supply chain}

\subparagraph{Value chain} The value chains concentrate on the differentiators related to market positioning within an organization and the cost reduction efforts for improving effectiveness. The main driver for value chains within an organization is internal activities. The monetization business model for data is an area of current research \cite{faroukhi2020big}. Defining the return-of-investment for the data efforts is vital to enable AMI. The AMI initiatives related to these value chain activities should be part of the business practice aspect of the AMI disruption framework (see section~\ref{section:InternalBusiness}).

\subparagraph{Supply chain} The supply chains manage the network of suppliers, organizations, and distributors to deliver products and services to the customer. Therefore, the supply chain activities can benefit from AMI. One of the promises of the Industry 4.0 standard is to collect data related to the manufacturing activities of an organization. Furthermore, the analytics constructed from these data resources will also increase the quality of the supply chain \cite{awwad2018big}. The AMI initiatives related to these supply chain activities should be part of the external third party and customer aspect of the AMI disruption framework (see section~\ref{section:ExternalCustomers}).

\subsection{Ecosystem}

The ecosystems concentrate on external partnerships and business models for creating more value for the product portfolio. As a result, these ecosystems can suffer from bottlenecks. For example, for AI-related activities, the DataOps and MLOps capabilities of the underlying platform could be a bottleneck for experimentation during AMI initiatives and deployment of AI/ML Products. Therefore, it is essential to review positive and negative interactions with the ecosystem to concentrate on optimal cases for new developments.

\subparagraph{Platform} The AI/ML industry segment that designs, develops, implements, and deploys AI/ML Products requires a platform and a set of base AI/ML Products to attract users. The platform could be monetized separately from the AI/ML Products use cases. An open -or federated platform should also allow third parties to run complementary AI/ML Products. This self-servicing model avoids some of the challenges related to the development of these complementary AI/ML Products \cite{adner2010value}.

The cost of a platform-based ecosystem can be an essential enabler for optimizing its value. Often, an ecosystem rewards high-value users with a lower price. Additionally, high-value users with high subsidization costs should get a low subsidy. In order to sustain an ecosystem, the business model should guarantee the return-of-investment. Depending on the market, different models have been proposed \cite{bourne2015perspective}.

In today's business activities, an ecosystem can incorporate multiple platforms. However, multiple platforms in combination with a lack of standardization for data ingestion and AI/ML Product deployments can result in additional bottlenecks for achieving the optimal potential of the ecosystem \cite{zhang2021emergence}.

\section{Future application of the framework}

\subparagraph{Web 3.0} The future of AI/ML solutions depends on technological advances. The era of hyper-connected internet-of-things (IoT), higher levels of decentralization, expanding the use of cryptography, virtual reality, more personalization, the introduction of 5G, etc., are prominent trends that introduce a transition into the next generation of the internet: \emph{Web 3.0}. Many components of this framework will remain applicable. Nevertheless, the DataOps and MLOps solutions combined with the AI/ML Products could change dramatically. Therefore, new domains as Edge AI \cite{wang2019edge}, federated learning \cite{li2020federated}, and decentralized AI \cite{harris2019decentralized} are expected to become more prominent. In the following sections, some of the most disruptive technologies: edge computing and blockchain, will be introduced to illustrate that this framework for disruptive AMI remains relevant.

\subsection{Edge computing}

\subparagraph{Purpose of the edge} Edge computing aims to bring data processing at or as near as possible to the source of data generation \cite{van2017edge}. One of the main drivers of this computing paradigm is to effectively handle the increasing amount of data that is generated on the internet \cite{khan2019edge}; therefore, the edge will have a broad impact across ISPs, CSPs, etc. The main benefits of the edge are low latency, higher security, better resource management with a positive impact on ESG factors, etc.

\subparagraph{Current business applications} Edge computing is applied in multiple domains. Illustratively, manufacturers transitioning into Industry 4.0 include multi-year business strategies, including the edge. Some prominent applications are electric vehicles, smart cities, energy utility technologies, commercial unmanned aerial vehicles (UAVs), healthcare smart devices, etc. Therefore, the business plan, AMI initiative, and AI/ML Solutions enable new business opportunities. 

\subparagraph{Risk assessment} Many organizations enable microservices architectures, scaled platforms, incorporating cloud-based infrastructures, etc. It can be very disruptive to include the edge-cloud. Furthermore, multiple architectural decisions need to be made for using the edge. Some of these architectures still include security concerns.

\subsection{Blockchain}

\subparagraph{Purpose of blockchain} The aim of blockchain, or distributed ledger technology (DLT), is to automize transactions by computing a consensus that a transaction is valid \cite{salah2019blockchain}. The authorization of transactions performed by a network of independent users, implemented by a consensus protocol, instead of a third party, i.e., government, bank, etc. Blockchain was designed as a reaction to the Global Financial Crisis in 2008 to prevent excessive financial risk. Each transaction is cryptographically signed and verified by all mining nodes; all mining nodes record chained blocks representing all transactions, the so-called distributed ledger. Currently, a blockchain categorizes into (1) public: everyone can join the blockchain, (2) private: only authorized users can join, (3) consensus or federated: only authorized and pre-selected users can join, and (4) blockchain as a service (BaaS): cloud service providers are providing blockchain-related services \cite{salah2019blockchain,asante2021distributed}.
This computational paradigm is often used for cryptocurrencies, i.e., Bitcoin, Ether, etc. Currently, one of the most critical advances in blockchain innovation is to use of so-called \emph{smart contracts} or decentralized applications (dApps). Smart contracts are programs that govern transactions; often, smart contracts are implemented on the Ethereum blockchain programmed in the Solidity \cite{buterin2013block}. Additionally, smart contracts are emerging into a novel blockchain-enabled peer-to-peer financial system. Finally, non-fungible tokens (NFT) are digital assets often in the form of multimedia, i.e., images, videos, music, etc.; mainly available on the Ethereum blockchain \cite{wang2021non,rehman2021nfts}.

\subparagraph{Current business applications} Blockchain is applied in a variety of domains. Since the introduction of cryptocurrencies, finance has been going through different innovation initiatives related to blockchain. In finance, blockchains are often called decentralized payment systems (DPS). Nevertheless, most online payment systems are not providing the option to pay with a cryptocurrency. Often, eCommerce does not use DSPs because of the high volatility of the value of a cryptocurrency \cite{mita2019stablecoin}. An effort to overcome this volatility is by introducing Stablecoins. Furthermore, decentralized finance (DeFi) attempts to provide digitally financial primitives, i.e., trading, lending, derivatives, asset management, insurance, etc., programmed with smart contracts. Alternatively, healthcare data management and personalized medicine applications could benefit from blockchain technology, i.e., patients' electronic health records (EHR), medical equipment supply chain, etc. Additionally, the Health Insurance Portability and Accountability Act of 1996 (HIPAA) enforces regulatory requirements for high standards for authentication and record sharing requirements; these requirements could be implemented with DLT. Another domain of innovation related to DLT is supply chain management. As digital transformation seeks cost reduction opportunities and impacts more efficient production, supply chain optimization is often identified as an opportunity. In the last decade, the supply chain network of an organization and its suppliers has increased in complexity \cite{mizgier2013bottleneck}. Therefore, risk management and information gathering of the supply chain have become more critical in order to identify bottlenecks. Consequently, DLT has become an emergent technology for the supply chain \cite{asante2021distributed}. Finally, blockchain is also used in other domains, i.e., distributed data storage \cite{zichichi2020framework}, agriculture \cite{griffin2021three}, and many more. Therefore, blockchain and DLT have evolved a lot in recent years. Therefore, the business plan, AMI initiatives, and AI/ML Solutions must be reviewed in detail to support these business opportunities.

\subparagraph{Risk assessment} Blockchains have great potential to be a key technology for the future internet. Nevertheless, there is a general risk. This technology was designed with the intent to overcome risk and improve security. Unfortunately,  blockchain has introduced another set of security risks \cite{singh2021blockchain}, i.e., 50 \% attack on the consensus protocol, distributed denial of service (DDOS), private key attacks, sybil attacks, eclipse attacks, liveness attacks, double spending attack, etc. Additionally, smart contract software development could benefit from security coding standards and more mature design patterns to avoid vulnerabilities. Furthermore, blockchain solutions could suffer from energy consumption that could hurt the ESG factors in an organization \cite{sulkowski2021sustainability,de2021true}. If blockchain became a mainstream technology, it would have a problematic impact in our environment \cite{mora2018bitcoin,saingre2021measuring}. The loss of a private key will result in the loss of assets on the blockchain. The scalability and the number of transactions on a blockchain could be limiting factors for some applications. Similar to data privacy and AI/ML regulation, DLT and blockchain regulation for cryptocurrencies and DeFi could change the landscape of this technology. Also, NFTs have security vulnerabilities. Zero-knowledge proofs (ZKP) is a cryptographic handshake that verifies specific information by disclosing it within the DLT; therefore, it could improve the security of many different blockchain applications.


\section{Conclusion}

\subparagraph{Three parts of the framework} A framework for disruptive AMI includes three main parts: (1) a business plan, (2) the AMI, and (3) the AI/ML solution. A business plan specifies the strategic plan and the cost analysis. The strategic plan needs to review the strategies applied for the organization (see table~\ref{tab:strategic_plan}). The cost analysis will review the transaction cost, the adjustment cost, the customer's costs, and the cost for generating value. Finally, the business plan should emphasize the AMI and the AI/ML solution required to achieve the business goals.

\subparagraph{Impact of the framework} The AMI of an organization will need to review the journey for achieving the required performance for a novel AI/ML Product for improving a business process. Therefore, it is essential to define how the AI/ML solution will fit into an organization's value chain, supply chain, platform, and ecosystem. The value chain would define the internal function with the organization, i.e., the core technology that generates business value, human resources, product -and project management, finance, cyber security, etc. The supply chain should identify bottlenecks in the supplier and customer network. Additionally, the platform incorporates tools based upon MLOps and DataOps capabilities that support collaboration related to the AMI initiatives and deployment of existing AI/ML solutions. Finally, the organization's ecosystem will make the AI/ML Factory available to a community of users and partners. For this community to be effective and efficient, standardization for data and self-service procedures should be established.

\subparagraph{Execution of the framework} This framework can be developed gradually. Since AI/ML will disrupt an organization's business processes and practices, it will require the entire organization to evolve. This adaptation strategy of an organization to define the human and AI integration will be instrumental for the fourth industrial revolution. The primary purpose of this manuscript is to extend the current business administration frameworks for innovation for AI/ML. 

\section*{acknowledgements}

We want to thank Ignatius Anandappa [Corporate VP at Raytheon Technologies] for the feedback on the manuscript. 



\section*{legal statement}

This work has been performed in Wim Verleyen's personal capacity. Therefore, this work reflects Wim Verleyen's personal view not necessarily the view of Raytheon Technologies.





\bibliography{references}

\begin{thebibliography}{159}
\providecommand{\natexlab}[1]{#1}
\providecommand{\url}[1]{\texttt{#1}}
\providecommand{\urlprefix}{}

\bibitem[{Agrawal et~al.(2017)Agrawal, Ajay and Gans, Joshua and Goldfarb,
  Avi}]{agrawal2017expect}
Agrawal A, Gans J, Goldfarb A, What to expect from artificial intelligence.
\newblock MIT Sloan Management Review; 2017.

\bibitem[{Daugherty and Wilson(2018)Daugherty, Paul R and Wilson, H
  James}]{daugherty2018human}
Daugherty PR, Wilson HJ.
\newblock Human+ machine: Reimagining work in the age of AI.
\newblock Harvard Business Press; 2018.

\bibitem[{Amabile(2020)Amabile, Teresa M}]{amabile2020creativity}
Amabile TM.
\newblock Creativity, artificial intelligence, and a world of surprises.
\newblock Academy of Management Discoveries 2020;6(3):351--354.

\bibitem[{De~Cremer and Kasparov(2021)De Cremer, David and Kasparov,
  Garry}]{de2021ai}
De~Cremer D, Kasparov G.
\newblock AI should augment human intelligence, not replace it.
\newblock Harvard Business Review 2021;.

\bibitem[{Calzon(2021)Bernardita Calzon}]{top2021ai}
Calzon B, Top 10 Analytics And Business Intelligence Trends For 2022.
\newblock Business Intelligence; 2021.

\bibitem[{Gao et~al.(2018)Gao, Jianfeng and Galley, Michel and Li,
  Lihong}]{gao2018neural}
Gao J, Galley M, Li L.
\newblock Neural approaches to conversational ai.
\newblock In: The 41st International ACM SIGIR Conference on Research \&
  Development in Information Retrieval; 2018. p. 1371--1374.

\bibitem[{Vaswani et~al.(2017)Vaswani, Ashish and Shazeer, Noam and Parmar,
  Niki and Uszkoreit, Jakob and Jones, Llion and Gomez, Aidan N and Kaiser,
  {\L}ukasz and Polosukhin, Illia}]{vaswani2017attention}
Vaswani A, Shazeer N, Parmar N, Uszkoreit J, Jones L, Gomez AN, et~al.
\newblock Attention is all you need.
\newblock In: Advances in neural information processing systems; 2017. p.
  5998--6008.

\bibitem[{Tran and Ulissi(2018)Tran, Kevin and Ulissi, Zachary
  W}]{tran2018active}
Tran K, Ulissi ZW.
\newblock Active learning across intermetallics to guide discovery of
  electrocatalysts for CO 2 reduction and H 2 evolution.
\newblock Nature Catalysis 2018;1(9):696--703.

\bibitem[{Schuhmacher et~al.(2020)Schuhmacher, Alexander and Gatto, Alexander
  and Hinder, Markus and Kuss, Michael and Gassmann,
  Oliver}]{schuhmacher2020upside}
Schuhmacher A, Gatto A, Hinder M, Kuss M, Gassmann O.
\newblock The upside of being a digital pharma player.
\newblock Drug discovery today 2020;25(9):1569--1574.

\bibitem[{Wainberg et~al.(2018)Wainberg, Michael and Merico, Daniele and
  Delong, Andrew and Frey, Brendan J}]{wainberg2018deep}
Wainberg M, Merico D, Delong A, Frey BJ.
\newblock Deep learning in biomedicine.
\newblock Nature biotechnology 2018;36(9):829--838.

\bibitem[{Mamoshina et~al.(2016)Mamoshina, Polina and Vieira, Armando and
  Putin, Evgeny and Zhavoronkov, Alex}]{mamoshina2016applications}
Mamoshina P, Vieira A, Putin E, Zhavoronkov A.
\newblock Applications of deep learning in biomedicine.
\newblock Molecular pharmaceutics 2016;13(5):1445--1454.

\bibitem[{Stokes et~al.(2020)Stokes, Jonathan M and Yang, Kevin and Swanson,
  Kyle and Jin, Wengong and Cubillos-Ruiz, Andres and Donghia, Nina M and
  MacNair, Craig R and French, Shawn and Carfrae, Lindsey A and
  Bloom-Ackermann, Zohar and others}]{stokes2020deep}
Stokes JM, Yang K, Swanson K, Jin W, Cubillos-Ruiz A, Donghia NM, et~al.
\newblock A deep learning approach to antibiotic discovery.
\newblock Cell 2020;180(4):688--702.

\bibitem[{Wu et~al.(2019)Wu, Nan and Phang, Jason and Park, Jungkyu and Shen,
  Yiqiu and Huang, Zhe and Zorin, Masha and Jastrzebski, Stanis{\l}aw and
  F{\'e}vry, Thibault and Katsnelson, Joe and Kim, Eric and
  others}]{wu2019deep}
Wu N, Phang J, Park J, Shen Y, Huang Z, Zorin M, et~al.
\newblock Deep neural networks improve radiologists’ performance in breast
  cancer screening.
\newblock IEEE transactions on medical imaging 2019;39(4):1184--1194.

\bibitem[{McKinney et~al.(2020)McKinney, Scott Mayer and Sieniek, Marcin and
  Godbole, Varun and Godwin, Jonathan and Antropova, Natasha and Ashrafian,
  Hutan and Back, Trevor and Chesus, Mary and Corrado, Greg S and Darzi, Ara
  and others}]{mckinney2020international}
McKinney SM, Sieniek M, Godbole V, Godwin J, Antropova N, Ashrafian H, et~al.
\newblock International evaluation of an AI system for breast cancer screening.
\newblock Nature 2020;577(7788):89--94.

\bibitem[{Yang et~al.(2019)Yang, Kevin K and Wu, Zachary and Arnold, Frances
  H}]{yang2019machine}
Yang KK, Wu Z, Arnold FH.
\newblock Machine-learning-guided directed evolution for protein engineering.
\newblock Nature methods 2019;16(8):687--694.

\bibitem[{Jumper et~al.(2021)Jumper, John and Evans, Richard and Pritzel,
  Alexander and Green, Tim and Figurnov, Michael and Ronneberger, Olaf and
  Tunyasuvunakool, Kathryn and Bates, Russ and {\v{Z}}{\'\i}dek, Augustin and
  Potapenko, Anna and others}]{jumper2021highly}
Jumper J, Evans R, Pritzel A, Green T, Figurnov M, Ronneberger O, et~al.
\newblock Highly accurate protein structure prediction with AlphaFold.
\newblock Nature 2021;596(7873):583--589.

\bibitem[{Silver et~al.(2017)Silver, David and Schrittwieser, Julian and
  Simonyan, Karen and Antonoglou, Ioannis and Huang, Aja and Guez, Arthur and
  Hubert, Thomas and Baker, Lucas and Lai, Matthew and Bolton, Adrian and
  others}]{silver2017mastering}
Silver D, Schrittwieser J, Simonyan K, Antonoglou I, Huang A, Guez A, et~al.
\newblock Mastering the game of go without human knowledge.
\newblock nature 2017;550(7676):354--359.

\bibitem[{Kahng(2021)Kahng, Andrew B}]{kahng2021ai}
Kahng AB, AI system outperforms humans in designing floorplans for microchips.
\newblock Nature Publishing Group; 2021.

\bibitem[{Wu et~al.(2021)Wu, Jia and Li, Chao and Gensheimer, Michael and
  Padda, Sukhmani and Kato, Fumi and Shirato, Hiroki and Wei, Yiran and
  Sch{\"o}nlieb, Carola-Bibiane and Price, Stephen John and Jaffray, David and
  others}]{wu2021radiological}
Wu J, Li C, Gensheimer M, Padda S, Kato F, Shirato H, et~al.
\newblock Radiological tumour classification across imaging modality and
  histology.
\newblock Nature Machine Intelligence 2021;3(9):787--798.

\bibitem[{Ragot et~al.(2020)Ragot, Martin and Martin, Nicolas and Cojean,
  Salom{\'e}}]{ragot2020ai}
Ragot M, Martin N, Cojean S.
\newblock AI-generated vs. human artworks. a perception bias towards artificial
  intelligence?
\newblock In: Extended abstracts of the 2020 CHI conference on human factors in
  computing systems; 2020. p. 1--10.

\bibitem[{Chui et~al.(2018)Chui, Michael and Manyika, James and Miremadi,
  Mehdi}]{chui2018ai}
Chui M, Manyika J, Miremadi M.
\newblock What AI can and can’t do (yet) for your business.
\newblock McKinsey Quarterly 2018;1:97--108.

\bibitem[{Hagendorff and Wezel(2020)Hagendorff, Thilo and Wezel,
  Katharina}]{hagendorff202015}
Hagendorff T, Wezel K.
\newblock 15 challenges for AI: or what AI (currently) can’t do.
\newblock AI \& SOCIETY 2020;35(2):355--365.

\bibitem[{Davenport and Ronanki(2018)Davenport, Thomas H and Ronanki,
  Rajeev}]{davenport2018artificial}
Davenport TH, Ronanki R.
\newblock Artificial intelligence for the real world.
\newblock Harvard business review 2018;96(1):108--116.

\bibitem[{Ransbotham et~al.(2020)Ransbotham, S and Khodabandeh, S and Kiron, D
  and Candelon, F and Chu, M and LaFountain, B}]{ransbotham2020expanding}
Ransbotham S, Khodabandeh S, Kiron D, Candelon F, Chu M, LaFountain B.
\newblock Expanding AI’s impact with organizational learning.
\newblock MIT Sloan Management Review and Boston Consulting Group 2020;p.
  1--15.

\bibitem[{Robertson et~al.(2021)Robertson, James and Fossaceca, John M and
  Bennett, Kelly W}]{robertson2021cloud}
Robertson J, Fossaceca JM, Bennett KW.
\newblock A Cloud-Based Computing Framework for Artificial Intelligence
  Innovation in Support of Multidomain Operations.
\newblock IEEE Transactions on Engineering Management 2021;.

\bibitem[{McEnery(2019)Thornton McEnery}]{gigo2019}
McEnery T, Data Analytics: Garbage In, Garbage Out; 2019.
\newblock
  \urlprefix\url{https://dealbreaker.com/2017/01/data-analytics-garbage-in-garbage-out}.

\bibitem[{Ereth(2018)Ereth, Julian}]{ereth2018dataops}
Ereth J.
\newblock DataOps-Towards a Definition.
\newblock LWDA 2018;2191:104--112.

\bibitem[{Munappy et~al.(2020)Munappy, Aiswarya Raj and Mattos, David Issa and
  Bosch, Jan and Olsson, Helena Holmstr{\"o}m and Dakkak, Anas}]{munappy2020ad}
Munappy AR, Mattos DI, Bosch J, Olsson HH, Dakkak A.
\newblock From ad-hoc data analytics to dataops.
\newblock In: Proceedings of the International Conference on Software and
  System Processes; 2020. p. 165--174.

\bibitem[{Tamburri(2020)Tamburri, Damian A}]{tamburri2020sustainable}
Tamburri DA.
\newblock Sustainable MLOps: Trends and Challenges.
\newblock In: 2020 22nd International Symposium on Symbolic and Numeric
  Algorithms for Scientific Computing (SYNASC) IEEE; 2020. p. 17--23.

\bibitem[{M{\"a}kinen et~al.(2021)M{\"a}kinen, Sasu and Skogstr{\"o}m, Henrik
  and Laaksonen, Eero and Mikkonen, Tommi}]{makinen2021needs}
M{\"a}kinen S, Skogstr{\"o}m H, Laaksonen E, Mikkonen T.
\newblock Who Needs MLOps: What Data Scientists Seek to Accomplish and How Can
  MLOps Help?
\newblock arXiv preprint arXiv:210308942 2021;.

\bibitem[{Highsmith(2009)Highsmith, Jim}]{highsmith2009agile}
Highsmith J.
\newblock Agile project management: creating innovative products.
\newblock Pearson education; 2009.

\bibitem[{Lakshmanan et~al.(2020)Lakshmanan, Valliappa and Robinson, Sara and
  Munn, Michael}]{lakshmanan2020machine}
Lakshmanan V, Robinson S, Munn M.
\newblock Machine learning design patterns.
\newblock O'Reilly Media; 2020.

\bibitem[{Gamma et~al.(1995)Gamma, Erich and Helm, Richard and Johnson, Ralph
  and Vlissides, John and Patterns, Design}]{gamma1995elements}
Gamma E, Helm R, Johnson R, Vlissides J, Patterns D.
\newblock Elements of reusable object-oriented software, vol.~99.
\newblock Addison-Wesley Reading, Massachusetts; 1995.

\bibitem[{Price and Meyers(2006)Price, Courtney and Meyers, Arlen
  D}]{price200612}
Price C, Meyers AD.
\newblock The 12-step innovation roadmap: how to analyze and prioritize new
  business ideas.
\newblock Physician executive 2006;32(2):52.

\bibitem[{Kijkuit and Van Den~Ende(2007)Kijkuit, Bob and Van Den Ende,
  Jan}]{kijkuit2007organizational}
Kijkuit B, Van Den~Ende J.
\newblock The organizational life of an idea: Integrating social network,
  creativity and decision-making perspectives.
\newblock Journal of Management Studies 2007;44(6):863--882.

\bibitem[{Osterwalder and Pigneur(2010)Osterwalder, Alexander and Pigneur,
  Yves}]{osterwalder2010business}
Osterwalder A, Pigneur Y.
\newblock Business model generation: a handbook for visionaries, game changers,
  and challengers, vol.~1.
\newblock John Wiley \& Sons; 2010.

\bibitem[{Blank(2013)Blank, Steve}]{blank2013lean}
Blank S.
\newblock Why the lean start-up changes everything.
\newblock Harvard business review 2013;91(5):63--72.

\bibitem[{Henderson and Clark(1990)Henderson, Rebecca M and Clark, Kim
  B}]{henderson1990architectural}
Henderson RM, Clark KB.
\newblock Architectural innovation: The reconfiguration of existing product
  technologies and the failure of established firms.
\newblock Administrative science quarterly 1990;p. 9--30.

\bibitem[{Richter et~al.(2018)Richter, Nancy and Jackson, Paul and Schildhauer,
  Thomas}]{richter2018entrepreneurial}
Richter N, Jackson P, Schildhauer T.
\newblock Entrepreneurial behaviour and startups: The case of Germany and the
  USA.
\newblock In: Entrepreneurial Innovation and Leadership Springer; 2018.p.
  1--14.

\bibitem[{Bower and Christensen(1996)Bower, Joseph L and Christensen, Clayton
  M}]{bower1996disruptive}
Bower JL, Christensen CM.
\newblock Disruptive technologies: Catching the wave.
\newblock The Journal of Product Innovation Management 1996;1(13):75--76.

\bibitem[{Tripsas and Gavetti(2000)Tripsas, Mary and Gavetti,
  Giovanni}]{tripsas2000capabilities}
Tripsas M, Gavetti G.
\newblock Capabilities, cognition, and inertia: Evidence from digital imaging.
\newblock Strategic management journal 2000;21(10-11):1147--1161.

\bibitem[{Ahuja and Morris~Lampert(2001)Ahuja, Gautam and Morris Lampert,
  Curba}]{ahuja2001entrepreneurship}
Ahuja G, Morris~Lampert C.
\newblock Entrepreneurship in the large corporation: A longitudinal study of
  how established firms create breakthrough inventions.
\newblock Strategic management journal 2001;22(6-7):521--543.

\bibitem[{Christensen(2013)Christensen, Clayton M}]{christensen2013innovator}
Christensen CM.
\newblock The innovator's dilemma: when new technologies cause great firms to
  fail.
\newblock Harvard Business Review Press; 2013.

\bibitem[{Blank(2016)Blank, Steve}]{blank2016}
Blank S, Moving Companies at Startup Speeds; 2016.
\newblock \urlprefix\url{https://www.youtube.com/watch?v=AX0EnxpkZ6I}.

\bibitem[{Baghai et~al.(2000)Baghai, Mehrdad and Coley, Stephen and White,
  David}]{baghai2000alchemy}
Baghai M, Coley S, White D.
\newblock The alchemy of growth.
\newblock Basic Books; 2000.

\bibitem[{Wu(2022)Lynn Wu}]{wu2022}
Wu L, How to Be Smart About Artificial Intelligence; 2022.
\newblock
  \urlprefix\url{https://magazine.wharton.upenn.edu/issues/fall-winter-2021/how-to-be-smart-about-artificial-intelligence/}.

\bibitem[{Foster(1986)Foster, Richard}]{foster1986innovation}
Foster R, Innovation--the attacker’s advantage. McKinsey and Co. Inc.
\newblock New York: Summit Books, A Division of Simon and Schuster; 1986.

\bibitem[{Ries(2014)Ries, Eric}]{ries2014lean}
Ries E.
\newblock Lean Startup: Establish a company quickly, risk-free and
  successfully.
\newblock Redline Economics; 2014.

\bibitem[{Farmer(2021)John Paul Farmer}]{nyc2021aistrategy}
Farmer JP.
\newblock AI Strategy.
\newblock The City of New York); 2021.

\bibitem[{Christensen and Overdorf(2000)Christensen, Clayton M and Overdorf,
  Michael}]{christensen2000meeting}
Christensen CM, Overdorf M.
\newblock Meeting the challenge of disruptive change.
\newblock Harvard business review 2000;78(2):66--77.

\bibitem[{Borchert and Hagenhoff(2003)Borchert, Jan Eric and Hagenhoff,
  Svenja}]{borchert2003operatives}
Borchert JE, Hagenhoff S.
\newblock Operatives Innovations-und Technologiemanagement: Eine
  Bestandsaufnahme.
\newblock Arbeitsbericht 2003;.

\bibitem[{Duncan(1976)Duncan, Robert B}]{duncan1976ambidextrous}
Duncan RB.
\newblock The ambidextrous organization: Designing dual structures for
  innovation.
\newblock The management of organization 1976;1(1):167--188.

\bibitem[{March(1991)March, James G}]{march1991exploration}
March JG.
\newblock Exploration and exploitation in organizational learning.
\newblock Organization science 1991;2(1):71--87.

\bibitem[{Tushman and O'Reilly~III(1996)Tushman, Michael L and O'Reilly III,
  Charles A}]{tushman1996ambidextrous}
Tushman ML, O'Reilly~III CA.
\newblock Ambidextrous organizations: Managing evolutionary and revolutionary
  change.
\newblock California management review 1996;38(4):8--29.

\bibitem[{Birkinshaw and Gibson(2004)Birkinshaw, Julian and Gibson,
  Cristina}]{birkinshaw2004building}
Birkinshaw J, Gibson C.
\newblock Building ambidexterity into an organization.
\newblock MIT Sloan management review 2004;45(4).

\bibitem[{Strategy(1979)Strategy, How Competitive Forces
  Shape}]{strategy1979michael}
Strategy HCFS.
\newblock by Michael E. Porter.
\newblock Harvard Business Review 1979;.

\bibitem[{Arent et~al.(2016)Arent, Douglas J and Pless, Jacquelyn and Statwick,
  Patricia}]{arent2016five}
Arent DJ, Pless J, Statwick P.
\newblock Five Forces of 21st Century Innovation Strategy: Insights for
  Leaders.
\newblock National Renewable Energy Lab.(NREL), Golden, CO (United States);
  2016.

\bibitem[{Isabelle et~al.(2020)Isabelle, Diane and Horak, Kevin and McKinnon,
  Sarah and Palumbo, Chiara}]{isabelle2020porter}
Isabelle D, Horak K, McKinnon S, Palumbo C.
\newblock Is Porter's Five Forces Framework Still Relevant? A study of the
  capital/labour intensity continuum via mining and IT industries.
\newblock Technology Innovation Management Review 2020;10(6).

\bibitem[{Sull et~al.(2022)Sull, Donald and Sull, Charles and and Zweig,
  Ben}]{sull2022innovation}
Sull D, Sull C, , Zweig B, Toxic Culture Is Driving the Great Resignation.
\newblock MIT Sloan Management Review; 2022.

\bibitem[{Kim(2020)Jane J. Kim}]{kim2020}
Kim JJ, Preparing for the Next 'Black Swan'; 2020.
\newblock
  \urlprefix\url{https://www.wsj.com/articles/SB10001424052748703791804575439562361453200}.

\bibitem[{Gailly(2018)Gailly, Benoit}]{gailly2018navigating}
Gailly B.
\newblock Navigating Innovation: How to Identify, Prioritize and Capture
  Opportunities for Strategic Success.
\newblock Springer; 2018.

\bibitem[{Sahal(1981)Sahal, Devendra}]{sahal1981patterns}
Sahal D.
\newblock Patterns of technological innovation, vol. 198.
\newblock Addison-Wesley Reading, MA; 1981.

\bibitem[{Brown(1992)Brown, Rick}]{brown1992managing}
Brown R.
\newblock Managing the “S” curves of innovation.
\newblock Journal of Business \& Industrial Marketing 1992;.

\bibitem[{Christensen(1992)Christensen, Clayton M}]{christensen1992exploring}
Christensen CM.
\newblock Exploring the limits of the technology S-curve. Part I: component
  technologies.
\newblock Production and operations management 1992;1(4):334--357.

\bibitem[{Tzinis(2021)Irene Tzinis}]{Tzinis2021}
Tzinis I, Technology Readiness Level; 2021.
\newblock
  \urlprefix\url{https://www.nasa.gov/directorates/heo/scan/engineering/technology/technology_readiness_level}.

\bibitem[{Wikipedia(2021)}]{wiki2021}
Wikipedia, Technology Readiness Level; 2021.
\newblock
  \urlprefix\url{https://en.wikipedia.org/wiki/Technology_readiness_level#cite_note-NASA_to_EU-1}.

\bibitem[{Fowler(2004)Fowler, Kim}]{fowler2004build}
Fowler K.
\newblock Build versus buy.
\newblock IEEE Instrumentation \& Measurement Magazine 2004;7(3):67--73.

\bibitem[{Constantiou and Kallinikos(2015)Constantiou, Ioanna D and Kallinikos,
  Jannis}]{constantiou2015new}
Constantiou ID, Kallinikos J.
\newblock New games, new rules: big data and the changing context of strategy.
\newblock Journal of Information Technology 2015;30(1):44--57.

\bibitem[{DalleMule and Davenport(2017)DalleMule, Leandro and Davenport, Thomas
  H}]{dallemule2017s}
DalleMule L, Davenport TH.
\newblock What’s your data strategy.
\newblock Harvard Business Review 2017;95(3):112--121.

\bibitem[{Bhattacharya(2021)Joydeep Bhattacharya}]{Bhattacharya2021}
Bhattacharya J, Scaling Data Science And AI To Boost Business Growth; 2021.
\newblock
  \urlprefix\url{https://www.datasciencecentral.com/profiles/blogs/scaling-data-science-and-ai-to-boost-business-growth}.

\bibitem[{Kiron and Schrage(2019)Kiron, David and Schrage,
  Michael}]{kiron2019strategy}
Kiron D, Schrage M.
\newblock Strategy for and with AI.
\newblock MIT Sloan Management Review 2019;60(4):30--35.

\bibitem[{Chuprina(2020)Roman Chuprina}]{Chuprina2020}
Chuprina R, The Complete Guide on Customer Demand Forecasting in Retail; 2020.
\newblock
  \urlprefix\url{https://www.datasciencecentral.com/profiles/blogs/the-complete-guide-on-customer-demand-forecasting-in-retail}.

\bibitem[{Northcote et~al.(2021)Mulligan Nicholas Northcote and Tido Rode and
  Sasha Vesuvala}]{Mulligan2021}
Northcote MN, Rode T, Vesuvala S, The strategy-analytics revolution; 2021.
\newblock
  \urlprefix\url{https://www.mckinsey.com/business-functions/strategy-and-corporate-finance/our-insights/the-strategy-analytics-revolution}.

\bibitem[{Vesuvala and Brown(2021)Sasha Vesuvala and Sean Brown}]{Vesuvala2021}
Vesuvala S, Brown S, Improving strategic outcomes with advanced analytics;
  2021.
\newblock
  \urlprefix\url{https://www.mckinsey.com/business-functions/strategy-and-corporate-finance/our-insights/improving-strategic-outcomes-with-advanced-analytics}.

\bibitem[{Van~Duuren et~al.(2016)Van Duuren, Emiel and Plantinga, Auke and
  Scholtens, Bert}]{van2016esg}
Van~Duuren E, Plantinga A, Scholtens B.
\newblock ESG integration and the investment management process: Fundamental
  investing reinvented.
\newblock Journal of Business Ethics 2016;138(3):525--533.

\bibitem[{Chen and Mussalli(2020)Chen, Mike and Mussalli,
  George}]{chen2020integrated}
Chen M, Mussalli G.
\newblock An integrated approach to quantitative ESG investing.
\newblock The Journal of Portfolio Management 2020;46(3):65--74.

\bibitem[{S{\ae}tra(2021)S{\ae}tra, Henrik Skaug}]{saetra2021framework}
S{\ae}tra HS.
\newblock A Framework for Evaluating and Disclosing the ESG Related Impacts of
  AI with the SDGs.
\newblock Sustainability 2021;13(15):8503.

\bibitem[{Liu et~al.(2021)Liu, Xinlai and Wu, Haoye and Wu, Wei and Fu, Yelin
  and Huang, George Q}]{liu2021blockchain}
Liu X, Wu H, Wu W, Fu Y, Huang GQ.
\newblock Blockchain-enabled ESG reporting framework for sustainable supply
  chain.
\newblock In: Sustainable Design and Manufacturing 2020 Springer; 2021.p.
  403--413.

\bibitem[{Lai(2021)Lai, Karry}]{lai2021blockchain}
Lai K.
\newblock How blockchain can help drive sustainable finance.
\newblock International Financial Law Review 2021;.

\bibitem[{Haefner et~al.(2021)Haefner, Naomi and Wincent, Joakim and Parida,
  Vinit and Gassmann, Oliver}]{haefner2021artificial}
Haefner N, Wincent J, Parida V, Gassmann O.
\newblock Artificial intelligence and innovation management: A review,
  framework, and research agenda.
\newblock Technological Forecasting and Social Change 2021;162:120392.

\bibitem[{Van Der~Aalst(2012)Van Der Aalst, Wil}]{van2012process}
Van Der~Aalst W.
\newblock Process mining.
\newblock Communications of the ACM 2012;55(8):76--83.

\bibitem[{Leno et~al.(2021)Leno, Volodymyr and Polyvyanyy, Artem and Dumas,
  Marlon and La Rosa, Marcello and Maggi, Fabrizio}]{leno2021robotic}
Leno V, Polyvyanyy A, Dumas M, La~Rosa M, Maggi F.
\newblock Robotic process mining: vision and challenges.
\newblock Business \& Information Systems Engineering 2021;63(3):301--314.

\bibitem[{Laura(2021)Laura, Trautmann}]{laura2021product}
Laura T.
\newblock Product customization and generative design.
\newblock Multidiszciplin{\'a}ris Tudom{\'a}nyok 2021;11(4):87--95.

\bibitem[{Basole and Accenture(2021)Basole, Rahul and Accenture,
  AI}]{basole2021visualizing}
Basole R, Accenture A.
\newblock Visualizing the Evolution of the AI Ecosystem.
\newblock In: HICSS; 2021. p. 1--10.

\bibitem[{Mauborgne and Furman(2006)Mauborgne, Ren{\'e}e and Furman,
  Z}]{mauborgne2006blue}
Mauborgne R, Furman Z.
\newblock Blue ocean strategy.
\newblock Recorded Books, Incorporated; 2006.

\bibitem[{Agnihotri(2016)Agnihotri, Arpita}]{agnihotri2016extending}
Agnihotri A.
\newblock Extending boundaries of blue ocean strategy.
\newblock Journal of Strategic Marketing 2016;24(6):519--528.

\bibitem[{Gans and Stern(2003)Gans, Joshua S and Stern,
  Scott}]{gans2003product}
Gans JS, Stern S.
\newblock The product market and the market for “ideas”: commercialization
  strategies for technology entrepreneurs.
\newblock Research policy 2003;32(2):333--350.

\bibitem[{Coase(1937)Coase, Ronald Harry}]{coase1937nature}
Coase RH.
\newblock The nature of the firm.
\newblock economica 1937;4(16):386--405.

\bibitem[{Williamson(1979)Williamson, Oliver E}]{williamson1979transaction}
Williamson OE.
\newblock Transaction-cost economics: the governance of contractual relations.
\newblock The journal of Law and Economics 1979;22(2):233--261.

\bibitem[{Cuypers et~al.(2021)Cuypers, Ilya RP and Hennart, Jean-Fran{\c{c}}ois
  and Silverman, Brian S and Ertug, Gokhan}]{cuypers2021transaction}
Cuypers IR, Hennart JF, Silverman BS, Ertug G.
\newblock Transaction cost theory: Past progress, current challenges, and
  suggestions for the future.
\newblock Academy of Management Annals 2021;15(1):111--150.

\bibitem[{Lucas~Jr(1967)Lucas Jr, Robert E}]{lucas1967adjustment}
Lucas~Jr RE.
\newblock Adjustment costs and the theory of supply.
\newblock Journal of political economy 1967;75(4, Part 1):321--334.

\bibitem[{Basu(1987)Basu, Parantap}]{basu1987adjustment}
Basu P.
\newblock An adjustment cost model of asset pricing.
\newblock International Economic Review 1987;p. 609--621.

\bibitem[{Meghir et~al.(1996)Meghir, Costas and Ryan, Annette and Van Reenen,
  John}]{meghir1996job}
Meghir C, Ryan A, Van~Reenen J.
\newblock Job creation, technological innovation and adjustment costs: Evidence
  from a panel of British firms.
\newblock Annales d'Economie et de Statistique 1996;p. 255--274.

\bibitem[{Christensen et~al.(2018)Christensen, Clayton M and McDonald, Rory and
  Altman, Elizabeth J and Palmer, Jonathan E}]{christensen2018disruptive}
Christensen CM, McDonald R, Altman EJ, Palmer JE.
\newblock Disruptive innovation: An intellectual history and directions for
  future research.
\newblock Journal of Management Studies 2018;55(7):1043--1078.

\bibitem[{Pasquinelli(2019)Pasquinelli, Kevin M}]{pasquinelli2adapt}
Pasquinelli KM.
\newblock Adapt Your IP Strategy for Artificial Intelligence.
\newblock The Journal of Robotics, Artificial Intelligence \& Law 2019;2.

\bibitem[{Martins and Terblanche(2003)Martins, Ellen-Caroline and Terblanche,
  Fransie}]{martins2003building}
Martins EC, Terblanche F.
\newblock Building organisational culture that stimulates creativity and
  innovation.
\newblock European journal of innovation management 2003;.

\bibitem[{Dobni(2008)Dobni, C Brooke}]{dobni2008measuring}
Dobni CB.
\newblock Measuring innovation culture in organizations: The development of a
  generalized innovation culture construct using exploratory factor analysis.
\newblock European journal of innovation management 2008;.

\bibitem[{Adner and Lieberman(2021)Adner, Ron and Lieberman,
  Marvin}]{adner2021disruption}
Adner R, Lieberman M.
\newblock Disruption through complements.
\newblock Strategy Science 2021;6(1):91--109.

\bibitem[{Kim and Mauborgne(1997)Kim, W Chan and Mauborgne,
  Ren{\'e}e}]{kim1997value}
Kim WC, Mauborgne R.
\newblock Value innovation.
\newblock Havard Business Review 1997;1.

\bibitem[{Kim and Mauborgne(1999)Kim, W Chan and Mauborgne,
  Renee}]{kim1999strategy}
Kim WC, Mauborgne R.
\newblock Strategy, value innovation, and the knowledge economy.
\newblock MIT Sloan Management Review 1999;40(3):41.

\bibitem[{Chen et~al.(2016)Chen, Chung-Jen and Lin, Bou-Wen and Lin, Ya-Hui and
  Hsiao, Yung-Chang}]{chen2016ownership}
Chen CJ, Lin BW, Lin YH, Hsiao YC.
\newblock Ownership structure, independent board members and innovation
  performance: A contingency perspective.
\newblock Journal of Business Research 2016;69(9):3371--3379.

\bibitem[{Hill and Davis(2017)Hill, Linda A and Davis, George}]{hill2017board}
Hill LA, Davis G.
\newblock The board’s new innovation imperative.
\newblock Harvard Business Review 2017;95(6):102--109.

\bibitem[{Balsmeier et~al.(2017)Balsmeier, Benjamin and Fleming, Lee and Manso,
  Gustavo}]{balsmeier2017independent}
Balsmeier B, Fleming L, Manso G.
\newblock Independent boards and innovation.
\newblock Journal of Financial Economics 2017;123(3):536--557.

\bibitem[{Board(2019)Board, Defense Innovation}]{board2019ai}
Board DI, AI Principles: Recommendations on the Ethical Use of Artificial
  Intelligence by the Department of Defense: Supporting Document.
\newblock Defense Innovation Board, November; 2019.

\bibitem[{Terwiesch and Ulrich(2009)Terwiesch, Christian and Ulrich, Karl
  T}]{terwiesch2009innovation}
Terwiesch C, Ulrich KT.
\newblock Innovation tournaments: Creating and selecting exceptional
  opportunities.
\newblock Harvard Business Press; 2009.

\bibitem[{Wooten and Ulrich(2017)Wooten, Joel O and Ulrich, Karl
  T}]{wooten2017idea}
Wooten JO, Ulrich KT.
\newblock Idea generation and the role of feedback: Evidence from field
  experiments with innovation tournaments.
\newblock Production and Operations Management 2017;26(1):80--99.

\bibitem[{Yang et~al.(2021)Yang, Zimeng and Yan, Song and Lad, Abhimanyu and
  Liu, Xiaowei and Guo, Weiwei}]{yang2021cascaded}
Yang Z, Yan S, Lad A, Liu X, Guo W.
\newblock Cascaded Deep Neural Ranking Models in LinkedIn People Search.
\newblock In: Proceedings of the 30th ACM International Conference on
  Information \& Knowledge Management; 2021. p. 4312--4320.

\bibitem[{Porter(1996)Michael E. Porter}]{Porter1996}
Porter ME, What is Strategy?; 1996.
\newblock
  \urlprefix\url{http://www.rcmewhu.com/upload/file/20150528/20150528184258_9036.pdf}.

\bibitem[{Maital(2016)Shlomo Maital}]{Maital2016}
Maital S, Video Lecture: Tool 4 - Activity Map; 2016.
\newblock
  \urlprefix\url{https://www.coursera.org/lecture/startup-entrepreneurship-from-idea-to-startup/video-lecture-tool-4-activity-map-fhRsG}.

\bibitem[{Richardson(2005)Richardson, James E}]{richardson2005business}
Richardson JE.
\newblock The business model: an integrative framework for strategy execution.
\newblock Available at SSRN 932998 2005;.

\bibitem[{Schneckenberg et~al.(2021)Schneckenberg, Dirk and Matzler, Kurt and
  Spieth, Patrick}]{schneckenberg2021theorizing}
Schneckenberg D, Matzler K, Spieth P.
\newblock Theorizing business model innovation: an organizing framework of
  research dimensions and future perspectives.
\newblock R\&D Management 2021;.

\bibitem[{Ali and Anwar(2021)Ali, Bayad Jamal and Anwar,
  Govand}]{ali2021marketing}
Ali BJ, Anwar G.
\newblock Marketing Strategy: Pricing strategies and its influence on consumer
  purchasing decision.
\newblock Ali, BJ, \& Anwar, G(2021) Marketing Strategy: Pricing strategies and
  its influence on consumer purchasing decision International journal of Rural
  Development, Environment and Health Research 2021;5(2):26--39.

\bibitem[{Conick(2017)Conick, Hal}]{conick2017past}
Conick H.
\newblock The past, present and future of AI in marketing.
\newblock Marketing News 2017;51(1):26--35.

\bibitem[{Mariani et~al.(2021)Mariani, Marcello M and Perez-Vega, Rodrigo and
  Wirtz, Jochen}]{mariani2021ai}
Mariani MM, Perez-Vega R, Wirtz J.
\newblock AI in marketing, consumer research and psychology: A systematic
  literature review and research agenda.
\newblock Psychology \& Marketing 2021;.

\bibitem[{van Esch and Stewart~Black(2021)van Esch, Patrick and Stewart Black,
  J}]{van2021artificial}
van Esch P, Stewart~Black J.
\newblock Artificial intelligence (AI): revolutionizing digital marketing.
\newblock Australasian Marketing Journal 2021;29(3):199--203.

\bibitem[{Jia and Stan(2021)Jia, Peiyi and Stan, Ciprian}]{jia2021artificial}
Jia P, Stan C.
\newblock Artificial Intelligence Factory, Data Risk, and VCs’ Mediation: The
  Case of ByteDance, an AI-Powered Startup.
\newblock Journal of Risk and Financial Management 2021;14(5):203.

\bibitem[{Cubric(2020)Cubric, Marija}]{cubric2020drivers}
Cubric M.
\newblock Drivers, barriers and social considerations for AI adoption in
  business and management: A tertiary study.
\newblock Technology in Society 2020;62:101257.

\bibitem[{Chauhan and Kshetri(2021)Chauhan, Preeti S and Kshetri,
  Nir}]{chauhan20212021}
Chauhan PS, Kshetri N.
\newblock 2021 state of the practice in data privacy and security.
\newblock Computer 2021;54(08):125--132.

\bibitem[{Khashooei et~al.(2021)Khashooei, Behnam Asadi and Vasenev, Alexandr
  and Kocademir, Hasan Alper and Mathijssen,
  Roland}]{khashooei2021architecting}
Khashooei BA, Vasenev A, Kocademir HA, Mathijssen R.
\newblock Architecting System of Systems Solutions with Security and
  Data-Protection Principles.
\newblock In: 2021 16th International Conference of System of Systems
  Engineering (SoSE) IEEE; 2021. p. 43--48.

\bibitem[{Kosseff(2019)Kosseff, Jeff}]{kosseff2019twenty}
Kosseff J.
\newblock The twenty-six words that created the Internet.
\newblock Cornell University Press; 2019.

\bibitem[{Syroid et~al.(2021)Syroid, Tetiana L and Kaganovska, Tetiana Y and
  Shamraieva, Valentyna M and Perederii, Olexander S and Titov, Ievgen B and
  Varunts, Larysa D}]{syroid2021personal}
Syroid TL, Kaganovska TY, Shamraieva VM, Perederii OS, Titov IB, Varunts LD.
\newblock The personal data protection mechanism in the European Union.
\newblock International Journal of Computer Science \& Network Security
  2021;21(5):113--120.

\bibitem[{Douglas(2021)Douglas, Erika}]{douglas2021digital}
Douglas E.
\newblock Digital Crossroads: The Intersection of Competition Law and Data
  Privacy.
\newblock Available at SSRN 3880737 2021;.

\bibitem[{Institute(2021)Ponemon Institute}]{ibmsec2021}
Institute P, Cost of a Data Breach - Report 2021; 2021.
\newblock \urlprefix\url{https://www.ibm.com/security/data-breach}.

\bibitem[{Roberts et~al.(2021)Roberts, Huw and Cowls, Josh and Morley, Jessica
  and Taddeo, Mariarosaria and Wang, Vincent and Floridi,
  Luciano}]{roberts2021chinese}
Roberts H, Cowls J, Morley J, Taddeo M, Wang V, Floridi L.
\newblock The Chinese approach to artificial intelligence: an analysis of
  policy, ethics, and regulation.
\newblock AI \& SOCIETY 2021;36(1):59--77.

\bibitem[{Friedman(1975)Friedman, Milton}]{friedman1975there}
Friedman M.
\newblock There's no such thing as a free lunch.
\newblock Open Court LaSalle, IL; 1975.

\bibitem[{Smuha(2021)Smuha, Nathalie A}]{smuha2021race}
Smuha NA.
\newblock From a ‘race to AI’to a ‘race to AI regulation’: regulatory
  competition for artificial intelligence.
\newblock Law, Innovation and Technology 2021;13(1):57--84.

\bibitem[{Lessig(1991)Lawrence Lessig}]{lessig1991horse}
Lessig L, The law of the horse: what cyberlaw might teach; 1991.
\newblock \urlprefix\url{https://cyber.harvard.edu/works/lessig/finalhls.pdf}.

\bibitem[{Hansen and Weiskopf(2021)Hansen, Hans Krause and Weiskopf,
  Richard}]{hansen2021universalizing}
Hansen HK, Weiskopf R.
\newblock From Universalizing Transparency to the Interplay of Transparency
  Matrices: Critical insights from the emerging social credit system in China.
\newblock Organization Studies 2021;42(1):109--128.

\bibitem[{Moore(1993)Moore, James F}]{moore1993predators}
Moore JF.
\newblock Predators and prey: a new ecology of competition.
\newblock Harvard business review 1993;71(3):75--86.

\bibitem[{Sumbaly et~al.(2013)Sumbaly, Roshan and Kreps, Jay and Shah,
  Sam}]{sumbaly2013big}
Sumbaly R, Kreps J, Shah S.
\newblock The big data ecosystem at linkedin.
\newblock In: Proceedings of the 2013 ACM SIGMOD International Conference on
  Management of Data; 2013. p. 1125--1134.

\bibitem[{Tansley(1935)Tansley, Arthur G}]{tansley1935use}
Tansley AG.
\newblock The use and abuse of vegetational concepts and terms.
\newblock Ecology 1935;16(3):284--307.

\bibitem[{T et~al.(1990)Houghton J T and Jenkins G J and Ephraums J
  J}]{etde_6041139}
T HJ, J JG, J EJ, Climate change; 1990.

\bibitem[{of~Michigan(2008)University of Michigan}]{Michigan2008}
of~Michigan U, Ecological Communities: Networks of Interacting Species; 2008.
\newblock
  \urlprefix\url{https://globalchange.umich.edu/globalchange1/current/lectures/ecol_com/ecol_com.html}.

\bibitem[{Adner and Kapoor(2010)Adner, Ron and Kapoor, Rahul}]{adner2010value}
Adner R, Kapoor R.
\newblock Value creation in innovation ecosystems: How the structure of
  technological interdependence affects firm performance in new technology
  generations.
\newblock Strategic management journal 2010;31(3):306--333.

\bibitem[{Hamel and Zanini(2018)Hamel, Gary and Zanini, Michele}]{hamel2018end}
Hamel G, Zanini M.
\newblock The end of bureaucracy.
\newblock Harvard Business Review 2018;96(6):50--59.

\bibitem[{Adner and Kapoor(2016)Adner, Ron and Kapoor,
  Rahul}]{adner2016innovation}
Adner R, Kapoor R.
\newblock Innovation ecosystems and the pace of substitution: Re-examining
  technology S-curves.
\newblock Strategic management journal 2016;37(4):625--648.

\bibitem[{Faroukhi et~al.(2020)Faroukhi, Abou Zakaria and El Alaoui, Imane and
  Gahi, Youssef and Amine, Aouatif}]{faroukhi2020big}
Faroukhi AZ, El~Alaoui I, Gahi Y, Amine A.
\newblock Big data monetization throughout Big Data Value Chain: a
  comprehensive review.
\newblock Journal of Big Data 2020;7(1):1--22.

\bibitem[{Awwad et~al.(2018)Awwad, Mohamed and Kulkarni, Pranav and Bapna,
  Rachit and Marathe, Aniket}]{awwad2018big}
Awwad M, Kulkarni P, Bapna R, Marathe A.
\newblock Big data analytics in supply chain: a literature review.
\newblock In: Proceedings of the international conference on industrial
  engineering and operations management; 2018. p. 418--425.

\bibitem[{Bourne et~al.(2015)Bourne, Philip E and Lorsch, Jon R and Green, Eric
  D}]{bourne2015perspective}
Bourne PE, Lorsch JR, Green ED.
\newblock Perspective: Sustaining the big-data ecosystem.
\newblock Nature 2015;527(7576):S16--S17.

\bibitem[{Zhang and Williamson(2021)Zhang, Marina Yue and Williamson,
  Peter}]{zhang2021emergence}
Zhang MY, Williamson P.
\newblock The emergence of multiplatform ecosystems: insights from China's
  mobile payments system in overcoming bottlenecks to reach the mass market.
\newblock Technological Forecasting and Social Change 2021;173:121128.

\bibitem[{Wang et~al.(2019)Wang, Xiaofei and Han, Yiwen and Wang, Chenyang and
  Zhao, Qiyang and Chen, Xu and Chen, Min}]{wang2019edge}
Wang X, Han Y, Wang C, Zhao Q, Chen X, Chen M.
\newblock In-edge ai: Intelligentizing mobile edge computing, caching and
  communication by federated learning.
\newblock IEEE Network 2019;33(5):156--165.

\bibitem[{Li et~al.(2020)Li, Tian and Sahu, Anit Kumar and Talwalkar, Ameet and
  Smith, Virginia}]{li2020federated}
Li T, Sahu AK, Talwalkar A, Smith V.
\newblock Federated learning: Challenges, methods, and future directions.
\newblock IEEE Signal Processing Magazine 2020;37(3):50--60.

\bibitem[{Harris and Waggoner(2019)Harris, Justin D and Waggoner,
  Bo}]{harris2019decentralized}
Harris JD, Waggoner B.
\newblock Decentralized and collaborative AI on blockchain.
\newblock In: 2019 IEEE International Conference on Blockchain (Blockchain)
  IEEE; 2019. p. 368--375.

\bibitem[{van~der Meulen(2017)van der Meulen, Rob}]{van2017edge}
van~der Meulen R.
\newblock What edge computing means for infrastructure and operations leaders.
\newblock Gartner, online, available, www gartner com 2017;.

\bibitem[{Khan et~al.(2019)Khan, Wazir Zada and Ahmed, Ejaz and Hakak, Saqib
  and Yaqoob, Ibrar and Ahmed, Arif}]{khan2019edge}
Khan WZ, Ahmed E, Hakak S, Yaqoob I, Ahmed A.
\newblock Edge computing: A survey.
\newblock Future Generation Computer Systems 2019;97:219--235.

\bibitem[{Salah et~al.(2019)Salah, Khaled and Rehman, M Habib Ur and
  Nizamuddin, Nishara and Al-Fuqaha, Ala}]{salah2019blockchain}
Salah K, Rehman MHU, Nizamuddin N, Al-Fuqaha A.
\newblock Blockchain for AI: Review and open research challenges.
\newblock IEEE Access 2019;7:10127--10149.

\bibitem[{Asante et~al.(2021)Asante, Mary and Epiphaniou, Gregory and Maple,
  Carsten and Al-Khateeb, Haider and Bottarelli, Mirko and Ghafoor, Kayhan
  Zrar}]{asante2021distributed}
Asante M, Epiphaniou G, Maple C, Al-Khateeb H, Bottarelli M, Ghafoor KZ.
\newblock Distributed Ledger Technologies in Supply Chain Security Management:
  A Comprehensive Survey.
\newblock IEEE Transactions on Engineering Management 2021;.

\bibitem[{Buterin(2017)Vitalik Buterin}]{buterin2013block}
Buterin V, Ethereum Whitepaper; 2017.
\newblock \urlprefix\url{https://ethereum.org/en/whitepaper/}.

\bibitem[{Wang et~al.(2021)Wang, Qin and Li, Rujia and Wang, Qi and Chen,
  Shiping}]{wang2021non}
Wang Q, Li R, Wang Q, Chen S.
\newblock Non-fungible token (NFT): Overview, evaluation, opportunities and
  challenges.
\newblock arXiv preprint arXiv:210507447 2021;.

\bibitem[{Rehman et~al.(2021)Rehman, Wajiha and e Zainab, Hijab and Imran,
  Jaweria and Bawany, Narmeen Zakaria}]{rehman2021nfts}
Rehman W, e~Zainab H, Imran J, Bawany NZ.
\newblock NFTs: Applications and Challenges.
\newblock In: 2021 22nd International Arab Conference on Information Technology
  (ACIT) IEEE; 2021. p. 1--7.

\bibitem[{Mita et~al.(2019)Mita, Makiko and Ito, Kensuke and Ohsawa, Shohei and
  Tanaka, Hideyuki}]{mita2019stablecoin}
Mita M, Ito K, Ohsawa S, Tanaka H.
\newblock What is stablecoin?: A survey on price stabilization mechanisms for
  decentralized payment systems.
\newblock In: 2019 8th International Congress on Advanced Applied Informatics
  (IIAI-AAI) IEEE; 2019. p. 60--66.

\bibitem[{Mizgier et~al.(2013)Mizgier, Kamil J and J{\"u}ttner, Matthias P and
  Wagner, Stephan M}]{mizgier2013bottleneck}
Mizgier KJ, J{\"u}ttner MP, Wagner SM.
\newblock Bottleneck identification in supply chain networks.
\newblock International Journal of Production Research 2013;51(5):1477--1490.

\bibitem[{Zichichi et~al.(2020)Zichichi, Mirko and Ferretti, Stefano and
  D’angelo, Gabriele}]{zichichi2020framework}
Zichichi M, Ferretti S, D’angelo G.
\newblock A framework based on distributed ledger technologies for data
  management and services in intelligent transportation systems.
\newblock IEEE Access 2020;8:100384--100402.

\bibitem[{Griffin et~al.(2021)Griffin, Terry W and Harris, Keith D and Ward,
  Jason K and Goeringer, Paul and Richard, Jessica A}]{griffin2021three}
Griffin TW, Harris KD, Ward JK, Goeringer P, Richard JA.
\newblock Three digital agriculture problems in cotton solved by distributed
  ledger technology.
\newblock Applied Economic Perspectives and Policy 2021;.

\bibitem[{Singh et~al.(2021)Singh, Saurabh and Hosen, ASM Sanwar and Yoon,
  Byungun}]{singh2021blockchain}
Singh S, Hosen AS, Yoon B.
\newblock Blockchain security attacks, challenges, and solutions for the future
  distributed iot network.
\newblock IEEE Access 2021;9:13938--13959.

\bibitem[{Sulkowski(2021)Sulkowski, Adam J}]{sulkowski2021sustainability}
Sulkowski AJ.
\newblock Sustainability (or ESG) Reporting: Recent Developments and the
  Potential for Better, More Proactive Management Enabled by Blockchain.
\newblock More Proactive Management Enabled by Blockchain (October 23, 2021)
  2021;.

\bibitem[{de~Vries et~al.(2021)de Vries, Alex and Gallersd{\"o}rfer, Ulrich and
  Klaa{\ss}en, Lena and Stoll, Christian}]{de2021true}
de~Vries A, Gallersd{\"o}rfer U, Klaa{\ss}en L, Stoll C.
\newblock The true costs of digital currencies: Exploring impact beyond energy
  use.
\newblock One Earth 2021;4(6):786--789.

\bibitem[{Mora et~al.(2018)Mora, Camilo and Rollins, Randi L and Taladay, Katie
  and Kantar, Michael B and Chock, Mason K and Shimada, Mio and Franklin, Erik
  C}]{mora2018bitcoin}
Mora C, Rollins RL, Taladay K, Kantar MB, Chock MK, Shimada M, et~al.
\newblock Bitcoin emissions alone could push global warming above 2 C.
\newblock Nature Climate Change 2018;8(11):931--933.

\bibitem[{Saingre et~al.(2021)Saingre, Dimitri and Ledoux, Thomas and Menaud,
  Jean-Marc}]{saingre2021measuring}
Saingre D, Ledoux T, Menaud JM.
\newblock Measuring performances and footprint of blockchains with BCTMark: a
  case study on Ethereum smart contracts energy consumption.
\newblock Cluster Computing 2021;p. 1--19.

\end{thebibliography}



\end{document}